\documentclass[preprintnumbers,prd,showpacs,floatfix,superscriptaddress,nofootinbib,twocolumn, letterpaper]{revtex4-2}
\pdfoutput=1

\usepackage{longtable}
\usepackage{bm}
\usepackage{relsize}
\usepackage{amsfonts}
\usepackage{amsmath}
\usepackage{amssymb,epsf}
\usepackage{latexsym}
\usepackage{graphicx,epsfig}
\usepackage{amssymb}
\usepackage{float}
\usepackage{epstopdf}
\usepackage[colorlinks=true,citecolor=blue,linkcolor=red,urlcolor=black]{hyperref}
\usepackage{dcolumn}
\usepackage{psfrag}
\usepackage{wrapfig}
\usepackage{makeidx}
\usepackage{epsf}
\usepackage{color}
\usepackage{multirow}
\usepackage{mathtools}

\newcommand{\nota}[1]{\textcolor{red}{\sf{[Nota: #1]}} }

\begin{document}

\title{Shadows of boson and Proca stars with thin accretion disks}

\author{Jo\~{a}o Lu\'{i}s Rosa}
\email{joaoluis92@gmail.com}
\affiliation{Institute of Physics, University of Tartu, W. Ostwaldi 1, 50411 Tartu, Estonia}

\author{Diego Rubiera-Garcia} \email{drubiera@ucm.es}
\affiliation{Departamento de F\'isica Te\'orica and IPARCOS,
	Universidad Complutense de Madrid, E-28040 Madrid, Spain}

\date{\today}

\begin{abstract} 
In this work, we obtain the shadow images of spherically symmetric scalar boson and Proca stars using analytical fittings of  numerical solutions, when illuminated by a geometrically thin accretion disk. We chose a sample of four boson and four Proca stars with radii ranging from more compact configurations with $R\sim 9M$ to more dilute configurations with $R\sim 20M$, where $M$ is the total mass of the bosonic star. In these configurations, the absence of the light sphere (the locus of unstable bound geodesics) makes the optical appearance of these stars to be dominated by a single luminous ring enclosing a central brightness depression, and no further photon rings are available. We show that if one considers face-on observations and a disk model whose emission is truncated at some finite radius at which the luminosity attains its maximum value, both the size of the shadow, as well as the luminosity and depth of the bright region, are heavily influenced by the emission profile, with the choice of the type and parameters of the bosonic stars in our samples having a sub-dominant influence. These differences are nonetheless significantly magnified when one allows the accretion disk to extend close enough to the center of the star. Our results point out that even though bosonic stars are horizonless and do not have a photon sphere, some of them may be able to produce conventional black hole shadow-like images provided that their compactness is large enough, thus being potentially consistent with current and future observations.
\end{abstract}


\maketitle

\section{Introduction}\label{sec:intro}

The verification of the gravitational deflection of light by the Sun in 1919 according to the predictions of General Relativity (GR) \cite{Crispino:2019yew} supported the validity of the theory and triggered the beginning of a new era in the study of gravitational physics. Exactly one hundred years afterwards, this collective effort culminated in the observation of {\it shadows}, i.e., the silhouette of a compact enough body which, when illuminated by its own accretion disk, yields a central brightness depression whose boundary consist of a bright ring composed of those photons which have undergone a strong gravitational lensing  \cite{Luminet:1979nyg,Falcke:1999pj}. Indeed, the measurements of the Event Horizon Telescope (EHT) Collaboration, on the optical appearance created by the super-heated plasma surrounding the supermassive object at the center of the M87 galaxy \cite{Akiyama:2019cqa},  and of Sgr A* \cite{EHT2022PaperI} - the center of our one Milky Way -, are fully consistent with this picture. This finding is expected to have a tremendous impact in our chances to test the nature of ultra-compact objects and the strong-field regime of GR \cite{EventHorizonTelescope:2020qrl}, providing great opportunities both at theoretical and observational levels.

In the theoretical description of shadows, a prominent role is played by the notion of {\it critical curve(s)} \cite{Bozza:2002zj,Perlick:2021aok}. Such curves are present, for instance, in the space-time of a Kerr(-Newman) black hole, on which one finds a photon shell of unstable orbits corresponding to the light rays approaching asymptotically the critical points of the radial ``potential"  \cite{Cunha:2018acu,Hou:2022gge}. In absence of rotation, this shell degenerates into a single circular critical curve associated to an extrema of the effective potential (i.e., a photon sphere). A wide variety of studies have explored the connection of this theoretical notion with its associated observational signatures, see e.g. \cite{cunningham,Cardoso:2019dte,Gralla:2020srx,Cardoso:2021sip}. This is so because for every compact object subdued below its photon sphere, light rays can wind several times around it, producing a thin photon ring embedded in the main bright ring of radiation and terminating at the outer edge of a central brightness depression (the shadow) \cite{Gralla:2019xty}. The exact properties of the shadow (size and depth) as well as those of the infinite sequence of light rings potentially stacked in the photon ring and indexed by the number of (half-)orbits  (i.e. its locations and corresponding luminosities) are determined by a delicate interaction between the background geometry and the optical, geometrical, and emission features of the accretion disk surrounding the object \cite{Vincent:2022fwj}. Separating the contributions of the background geometry and the astrophysics of the disk illuminating it in shadow images has become a critical challenge in the field \cite{Lara:2021zth} in order to optimize the chances of finding new physics \cite{Wielgus:2021peu}, or in confirming the utter reliability of the Kerr solution to describe every astrophysical black hole \cite{Vincent:2020dij}.

Besides the Kerr black hole, at the theoretical level there are many other compact objects in the cosmic zoo with a wide range of properties, see e.g. \cite{Cardoso:2019rvt,Barack:2018yly} for reviews of these objects and their observational status. For the sake of their shadows, they may have a different description in the structure of their photon spheres, such as having more than one \cite{Wielgus:2020uqz,Tsukamoto:2021caq,Olmo:2021piq,Guerrero:2022qkh,Tsukamoto:2022vkt}, or none at all, which important observational consequences for their cast images. For instance, it has been recently proven that for asymptotically flat and horizonless exotic compact objects, if the space-time is topologically trivial, an even number of light-rings necessarily arise \cite{Cunha:2017qtt}. Here we are interested in the kind of objects that go collectively by the name of boson stars \cite{Vincent:2015xta,Olivares:2018abq,Guth:2014hsa,Macedo:2013jja,Macedo:2013qea,Brito:2015yfh,Cunha:2017wao,Herdeiro:2017fhv,Herdeiro:2019mbz,Annulli:2020lyc} (see \cite{Liebling:2012fv,Schunck:2003kk} for more extensive reviews), and which can be supported, for instance, by a massive complex scalar and vector (Proca) fields \cite{Brito:2015pxa,Minamitsuji:2018kof}, and which are typically not compact enough to hold a critical curve \cite{Cardoso:2021ehg}. The relevance of such solutions lies on the the fact that Proca stars can be able to mimic the shadow of a Schwarzschild black hole for low-inclination observations with a truncated accretion disk \cite{Herdeiro:2021lwl}, as well as orbital astrometric data \cite{Rosa:2022toh}, which emphasizes them as suitable alternatives to the black hole paradigm.

The main aim of this work is therefore to build shadow images of both these objects under geometrically thin accretion disks. The absence of a critical curve makes the photon (sub-)ring structure ascribed to the Kerr (Schwarzschild) solution to be missing for these objects, the corresponding images being instead completely dominated by the direct emission of the disk, and forming a single bright ring which bounds a central dark area. This work is then organized as follows: In Sec. \ref{sec:theory} we introduce the Einstein-Klein-Gordon and Einstein-Proca theories, of which the scalar boson and the Proca stars are solutions, respectively. We obtain four plus four samples of such spherically symmetric boson and Proca stars presented under analytical form, and corresponding to a fitting of numerical solutions. In Sec. \ref{sec:shadows} we assume five toy models for the accretion disk, with a monochromatic emission truncated at a finite radius (located at different relevant surfaces) where the luminosity takes its maximum value, and assuming different decays with the radial distance. We discuss how these two aspects of the modeling of the disk are the main force driving the shadow images (size of the black central area and luminosity and depth of the luminous ring) of these two objects, while the choice of the parameters of the solutions has a sub-dominant influence for the face-on observations considered. We also discuss how, for the particular case in which the inner edge of the disk approaches close enough to the center of the object, the characteristics of the bosonic star, i.e., its compactness and type of fundamental field, take a more active role in determining the features of the shadow.  Finally, in Sec. \ref{sec:concl}, we trace our conclusions on how future very-long baseline interferometry experiments could be able to distinguish, via the analysis (or the absence thereof) of higher-order light rings, between the Kerr black hole and any of its many kins proposed in the literature such as the boson and Proca stars studied here. 


\section{Theory and framework}\label{sec:theory}

\subsection{Einstein-Klein-Gordon theory}\label{sec:scalar}

The first type of bosonic star we are interested in modeling in this work is the scalar boson star. In this section, we shall mainly follow the analysis of e.g. Ref. \cite{Visinelli:2021uve}, suitably adapted to our needs. These stars are localized self-gravitating solutions of a massive and complex scalar field $\Phi$, which can be formulated within the Einstein-Klein-Gordon theory. This theory is described by an action  of the form
\begin{equation}\label{ekg_action}
\mathcal{S}=\int_\Omega \sqrt{-g}\left(\frac{R}{16\pi}-\nabla_a\bar\Phi\nabla^a\Phi-\mu^2\bar\Phi\Phi\right)d^4x ,
\end{equation}
where $\Omega$ is the spacetime manifold on which one defines a set of coordinates $x^a$, $g_{ab}$ is the metric tensor written in terms of the coordinates $x^a$, and whose determinant is denoted by $g$,  while $R=g^{ab}R_{ab}$ is the Ricci scalar, with $R_{ab}$ the Ricci tensor. As for the scalar field $\Phi$, the object $\nabla_a$ denotes covariant derivatives written in terms of the metric $g_{ab}$, an overbar ($\bar{\ }$) denotes complex conjugation, and $\mu$ is a parameter representing the mass of the scalar field. In our conventions, we have adopted a system of geometrized units on which $G=c=1$, where $G$ is the gravitational constant and $c$ is the speed of light. Note also that Eq.\eqref{ekg_action} represents the simplest possible form of the Einstein-Klein-Gordon theory with a massive scalar field, but more complicated forms of the theory could be considered via the addition of higher-order interaction terms (e.g. quartic terms) in $\Phi$, though for the sake of this work we shall take this simplified form of the action. 

The equilibrium solutions describing scalar boson stars can be obtained by solving the equations of motion of the theory. Since the action in Eq. \eqref{ekg_action} depends explicitly on two quantities, namely the metric $g_{ab}$ and the scalar field $\Phi$, taking its variation with respect to these two fields leads to the so-called Einstein-Klein-Gordon system described by
\begin{eqnarray}
R_{ab}-\frac{1}{2}g_{ab}R&=&8\pi T_{ab}, \label{ekg_field} \\ 
\left(\Box-\mu^2\right)\Phi&=&0, \label{ekg_eomphi}
\end{eqnarray}
where $\Box=\nabla^a\nabla_a$ represents the d'Alembert operator and the stress-energy tensor $T_{ab}$ is given in terms of the scalar field $\Phi$ and its derivatives as
\begin{equation}\label{ekg_tab}
T_{ab}=\nabla_{(a}\bar\Phi\nabla_{b)}\Phi-\frac{1}{2}g_{ab}\left(\nabla_c\bar\Phi\nabla^c\Phi+\mu^2\bar\Phi\Phi\right),
\end{equation}
where the parenthesis $X_{(ab)}=\left(X_{ab}+X_{ba}\right)/2$ denotes index symmetrization. For simplicity, we will focus exclusively in static and spherically symmetric solutions of the Einstein-Klein-Gordon system. These solutions can be generically described by a line element in the usual spherical coordinate system $\left(t,r,\theta,\phi\right)$ of the form
\begin{equation}\label{ekg_metric}
ds^2=-\sigma^2(r)N(r)dt^2+\frac{1}{N(r)}dr^2+r^2d\Omega^2,
\end{equation}
where $\sigma(r)$ and $N(r)=1-2m(r)/r$ are metric functions assumed to depend exclusively on the radial coordinate $r$, with $m\left(r\right)$ a function that plays the role of the mass, and $d\Omega^2=d\theta^2+\sin^2\theta d\phi^2$ denotes the surface-element of the two-spheres. The time-independence of the metric $g_{ab}$ and the stress-energy tensor $T_{ab}$ can be preserved via the use of the $U(1)$ symmetry of the action in Eq. \eqref{ekg_action}, i.e., by considering a standing wave ansatz for the scalar field $\Phi$ as
\begin{equation}\label{ekg_scalar}
\Phi\left(t,r\right)=\frac{\gamma\left(r\right)}{\sqrt{8\pi}}e^{i\omega t},
\end{equation}
where $\gamma\left(r\right)$ represents the radial wave function and $\omega$ is a real constant representing the angular frequency of $\Phi$.

Under the assumptions described above, the Einstein-Klein-Gordon system of Eqs. \eqref{ekg_field} and \eqref{ekg_eomphi}, with $T_{ab}$ given by Eq. \eqref{ekg_tab} and the ansatze for the metric $g_{ab}$ and the scalar field $\Phi$ given respectively by Eqs. \eqref{ekg_metric} and \eqref{ekg_scalar}, becomes a system of three coupled ordinary differential equations (ODEs) for the metric functions $\sigma$ and $N$ and the radial wave function $\gamma$, forming the system
\begin{eqnarray}
N'&=&\frac{1}{r}-N\left(\frac{1}{r}+r\gamma'^2\right)-\frac{\omega^2r\gamma^2}{N\sigma^2}-\mu^2r\gamma^2\,,\\
\sigma'&=&r\frac{\omega^2\gamma^2+N^2\sigma^2\gamma'^2}{N^2\sigma}\,,\\ 
\gamma''&=&\frac{\gamma (\mu^2 rN\sigma^2-\omega^2 r)-N\sigma \gamma'\left(r\sigma N'+N(2\sigma+r\sigma')\right)}{rN^2\sigma^2}\,, 
\end{eqnarray}
To preserve its regularity at the origin $r=0$, i.e., to avoid the presence of any singular quantities, the functions $\sigma$, $N$ and $\gamma$ must satisfy some boundary conditions. These can be obtained by taking a series expansion at $r=0$ and are given by
\begin{eqnarray}
&m\left(r\right)=\mathcal O\left(r^3\right),\nonumber \\
&\sigma\left(r\right)=\sigma_0+\mathcal O\left(r^2\right), \label{ekg_boundary}\\
&\gamma\left(r\right)=\gamma_0+\mathcal O\left(r^2\right),\nonumber 
\end{eqnarray}
where $\sigma_0$ and $\gamma_0$ are constant parameters and $\mathcal O\left(r^n\right)$ collectively represents the arbitrariness in terms of order $n$ or higher in $r$. Furthermore, since we are interested in localized solutions of the scalar field $\Phi$ that preserve asymptotic flatness,  the wave function $\gamma$ is required to decay exponentially as $r\to \infty$, and the functions $\sigma$ and $N$ to approach unity in the same limit.

Given the complexity of the Einstein-Klein-Gordon system of equations, finding analytical solutions for bosonic star configurations is an unattainable task, and one usually resorts to numerical methods for the purpose. A common procedure is to consider a shooting method for the parameter $\omega$, considering a fixed combination of the remaining parameters $\mu$, $\sigma_0$ and $\gamma_0$ at the origin. Also, the parameter $\mu$ can be renormalized to unity with a redefinition of the radial coordinate in the form $x=\mu r$.

\subsection{Einstein-Proca theory}\label{sec:vector}

The second type of bosonic star we will be dealing with in this work is the vectorial boson star, or Proca star. In this section, we shall mainly review the formalism introduced in e.g. Ref. \cite{Brito:2015pxa}. These are localized solutions of a massive and complex vector field $A^a$, which are described by the Einstein-Proca theory. Its action is given by
\begin{equation}\label{ep_action}
\mathcal{S}=\int_\Omega  \sqrt{-g}\left(\frac{R}{16\pi}-\frac{1}{4}F_{ab}\bar F^{ab}-\frac{1}{2}\mu^2A_a \bar A^a\right)d^4x,
\end{equation}
where most of the quantities and notation have already been defined after Eq. \eqref{ekg_action}, while $F_{ab}=\partial_aA_b-\partial_bA_a$ is the electromagnetic tensor. Eq. \eqref{ep_action} is the simplest possible form of the action for an Einstein-Proca theory with a massive vector field, while likewise in the scalar field case more complicated theories could be exploited via the addition of higher-order interaction terms in the vector field.

Similarly to the scalar case introduced previously, the equilibrium configurations are obtained by solving the equations of motion of the system. In this case, these will be the field equations and the Proca equation, obtained from a variation of Eq. \eqref{ekg_action} with respect to the metric $g_{ab}$ and the vector field $A^a$, respectively. These equations form the so-called Einstein-Proca system and are given by
\begin{eqnarray}
R_{ab}-\frac{1}{2}g_{ab}R&=&8\pi T_{ab}, \label{ep_field} \\
\nabla_bF^{ab}&=&\mu^2A^a, \label{ep_eoma}
\end{eqnarray}
where the stress-energy tensor $T_{ab}$ is given in this case in terms of $A^a$ and its derivatives (contained within the tensor $F_{ab}$) as
\begin{equation}\label{ep_tab}
T_{ab}=-F_{c(a}\bar F^c_{\ b)}-\frac{1}{4}g_{ab}F_{cd}\bar F^{cd}+\mu^2\left[A_{(a}\bar A_ {b)}-\frac{1}{2}g_{ab}A_c\bar A^c\right].
\end{equation}
We again focus on static and spherically symmetric solutions of the Einstein-Proca system above. Again, we take the metric to be written in the generic form of Eq. \eqref{ekg_metric} where the metric functions $\sigma(r)$ and $N(r)$ remain functions of the radial coordinate $r$ only. The $U(1)$ symmetry of Eq. \eqref{ep_action} can again be exploited to preserve the staticity of the metric $g_{ab}$ and the stress-energy tensor $T_{ab}$, i.e., we impose a standing wave ansatz for the vector field as
\begin{equation}\label{ep_vector}
A_a=e^{-i\omega t}\left(f(r),i g(r),0,0\right),
\end{equation}
where the functions $f(r)$ and $g(r)$ are well-behaved functions of the radial coordinate only.

Under these assumptions, the Einstein-Proca system of Eqs. \eqref{ep_field} and \eqref{ep_eoma}, with a $T_{ab}$ given by Eq. \eqref{ep_tab} and ansatze for the metric $g_{ab}$ and the vector field $A^a$ in the forms of Eqs. \eqref{ekg_metric} and \eqref{ep_vector}, respectively, gives rise to four independent equations, two from the gravitational field equations, and two from the Proca equation. The field equations become
\begin{eqnarray}
m'&=&4\pi r^2\left[\frac{\left(f'-\omega g\right)^2}{2\sigma^2}+\frac{1}{2}\mu^2\left(g^2N+\frac{f^2}{N\sigma^2}\right)\right], \label{ep_field1} \\
\sigma'&=&4\pi r\mu^2\sigma\left(g^2+\frac{f^2}{N\sigma^2}\right), \label{ep_field2}
\end{eqnarray}
whereas the Proca equation splits into the system
\begin{eqnarray}
\left[\frac{r^2\left(f'-\omega g\right)}{\sigma}\right]'&=&\frac{\mu^2 r^2 f}{\sigma N}, \\ \label{ep_proca1}
\omega g-f'&=&\frac{\mu^2\sigma^2 Ng}{\omega}. \label{ep_proca2}
\end{eqnarray}
These equations form a system of four coupled ODEs for the functions $\sigma$, $N$, $f$, and $g$, which is again singular at the origin. To preserve the regularity of the system, a series expansion of the functions at $r=0$ requires the following boundary conditions to be satisfied
\begin{eqnarray}
f\left(r\right)&=&f_0+\mathcal O\left(r^2\right),\nonumber\\
g\left(r\right)&=&\mathcal O\left(r\right),\label{boundary}\\
m\left(r\right)&=&\mathcal O\left(r^3\right),\nonumber\\
\sigma\left(r\right)&=&\sigma_0+\mathcal O\left(r^2\right),\nonumber
\end{eqnarray}
where $f_0$ and $\sigma_0$ are constant parameters. Finally, since we are interested in localized solutions of the vector field $A^a$ that preserve asymptotic flatness, we also require that the functions $f$ and $g$ decay exponentially and the metric functions $N$ and $\sigma$ to approach unity in the limit $r\to \infty$.

To solve the system of Eqs. \eqref{ep_field1} to \eqref{ep_proca2}, one must again resort to numerical methods. A shooting method for $\omega$ while keeping the remaining parameters $\mu$, $\sigma_0$ and $f_0$ constant is the most common, and in particular the parameter $\mu$ can be renormalized to unity via a redefinition of the radial coordinate as $x=\mu r$.

\subsection{Solutions and approximations}\label{sec:solutions}

In the analysis that follows below, we shall consider eight specific bosonic star configurations, four scalar and four vectorial, obtained by numerically shooting-out the corresponding field equations under the conditions discussed in Secs. \ref{sec:scalar} and \ref{sec:vector}, respectively. The parameter used for the shooting method was the frequency $\omega$, which was computed to a precision of 12 decimal figures in the scalar boson star case and 15 decimal figures in the Proca star case, up to a numerical infinity set at $r/M=50$. Given the high precision of the numerical integration considered, the numerical errors of the solutions considered are negligible. Furthermore, to select an appropriate set of solutions, we have considered two criteria. Firstly, defining the binding energy of the bosonic star as $M-\mu Q$, where $Q$ is the conserved Noether charge, i.e., the number of particles, one verifies that there exists a value of the central densities $\gamma_0$ and $f_0$ for which the binding energy transitions from negative to positive, indicating an excess of energy. Our most compact configurations, namely BS1 and PS1, were chosen to stand on this transition, where the binding energy defined above is approximately zero. Secondly, taking into consideration the linear stability of bosonic star configurations, one verifies that both boson and Proca stars feature a maximum central density above which the configurations become linearly unstable. In the stability regime, the mass $\mu M$ of the star increases monotonically with the central density until the maximum density (and consequently, the maximum mass) are achieved. The maximum masses for scalar boson stars were found to be $\mu M\sim 0.633$\cite{Gleiser:1988rq}, whereas the maximum masses for Proca stars were found to be $\mu M\sim 1.06$\cite{Brito:2015pxa}. Our configurations BS2 and PS2 were chosen to stand in this transition, thus being marginally stable against linear perturbations, followed by the remaining configurations which are effectively linearly stable. Note however that this indicates that the configurations BS1 and PS1 stand in the unstable branch of bosonic star solutions. A summary of this analysis can be found in Ref. \cite{Cardoso:2019rvt}. Since in this work we focus on static bosonic star models with a single self-interaction potential, to preserve linear stability we have excluded from our analysis ultra-compact configurations, in order to avoid the potential instabilities associated to multiple critical curves and anti-photon spheres \cite{Cardoso:2014sna}. More interesting linearly stable solutions containing light rings could be considered via the addition of rotation or non-linear self interactions.

\begin{figure*}[t!]
\includegraphics[scale=0.85]{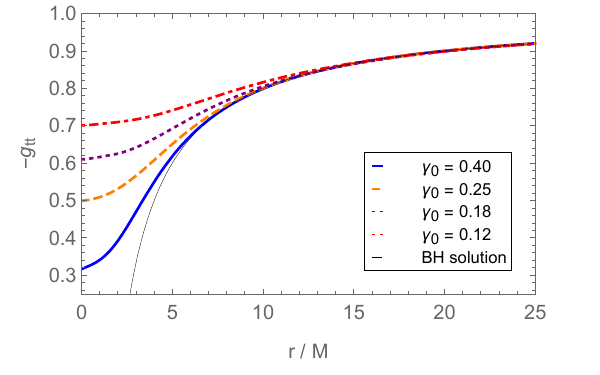}
\includegraphics[scale=0.85]{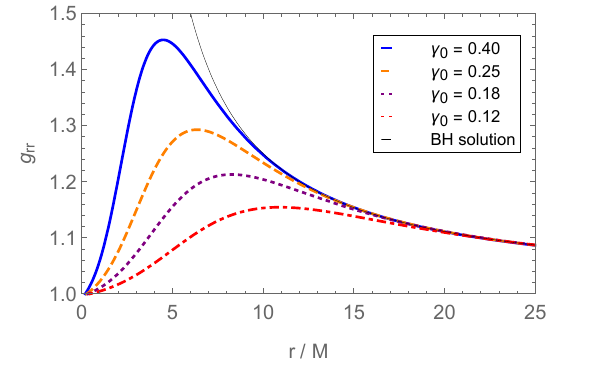}
\includegraphics[scale=0.85]{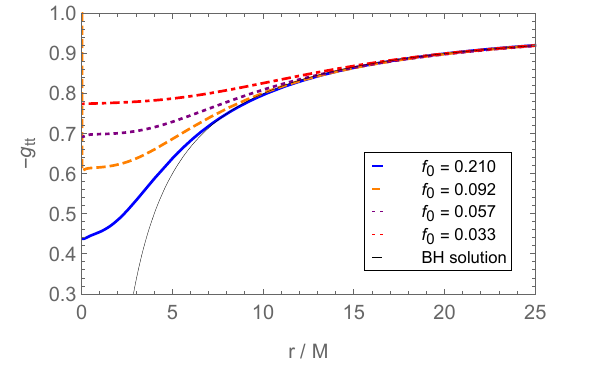}
\includegraphics[scale=0.85]{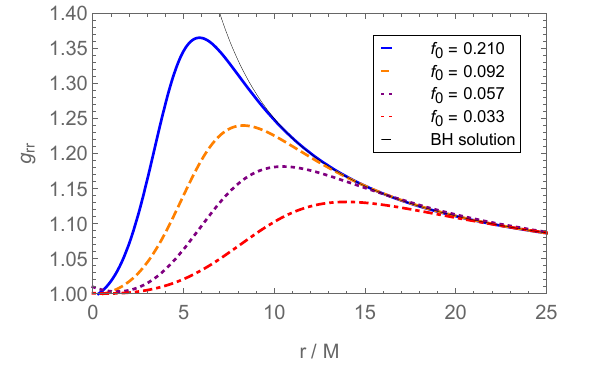}
\caption{The metric functions $-g_{tt}$ (left) and $g_{rr}$ (right) for the scalar boson stars (top) \{BS1,BS2,BS3,BS4\} (blue, orange, violet, red) in Table \ref{tab:bs_details}, and for the Proca stars  \{PS1,PS2,PS3,PS4\} (blue, orange, violet, red) in Table \ref{tab:ps_details}, both as a function of the normalized radial coordinate $r/M$.  The thin black line represents the Schwarzschild solution, i.e., $-g_{tt}=g_{rr}^{-1}=1-2M/r$.}
\label{fig:bs_metric}
\end{figure*}

\begin{figure*}[t!]
\includegraphics[scale=0.85]{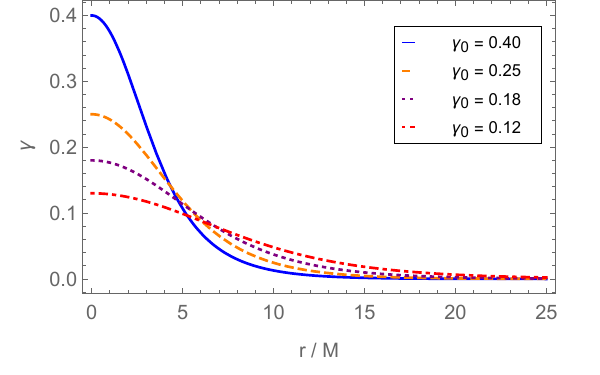}\\
\includegraphics[scale=0.85]{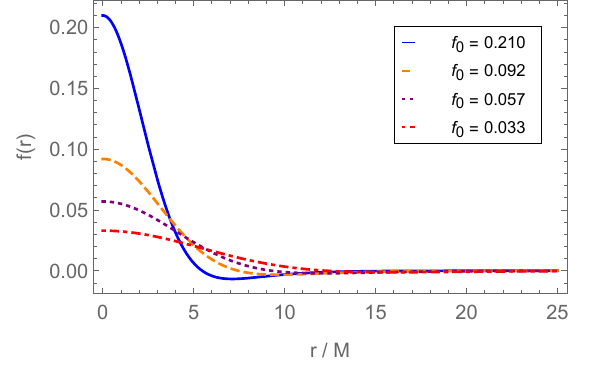}
\includegraphics[scale=0.85]{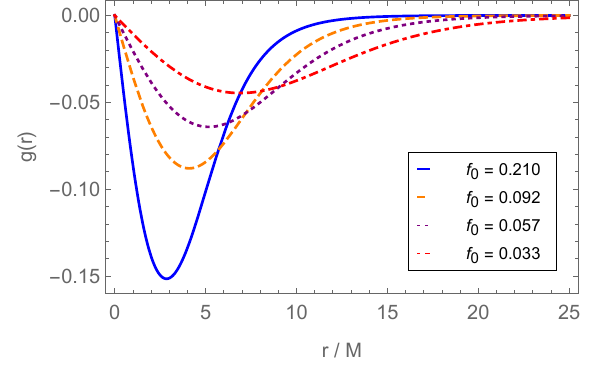}
\caption{Scalar field distribution $\gamma$ (top) and vector field distributions $f$ (bottom left) and $g$ (bottom right) for the scalar boson stars (top) \{BS1,BS2,BS3,BS4\} (blue, orange, violet, red) in Table \ref{tab:bs_details}, and for the Proca stars  \{PS1,PS2,PS3,PS4\} (blue, orange, violet, red) in Table \ref{tab:ps_details}, both as a function of the normalized radial coordinate $r/M$. The exponential decay at large radii confirm the localization of the fields.}
\label{fig:fields}
\end{figure*}

Therefore, the scalar boson stars employed in our analysis are dubbed as \{BS1,BS2,BS3,BS4\} and their characterizing parameters are detailed in Table \ref{tab:bs_details}, while the Proca stars are dubbed as \{PS1,PS2,PS3,PS4\} and are characterized in Table \ref{tab:ps_details}. These configurations differ not only in the type of fundamental field, i.e., scalar or vectorial, but also in their compactness, with radii varying from the most compact stable solutions with $R\sim 9M$ to more disperse solutions with $R\sim 20M$, where $R$ is defined as the radius encapsulating 98\% of the total mass of the star $M$, i.e., $m\left(R\right)=0.98M$, where the mass $M$ is defined as $M=\displaystyle{\lim_{r\to\infty}} m\left(r\right)$. The corresponding metric functions $g_{tt}$ and $g_{rr}$ (relevant for the analysis of shadows below) for both classes of configurations are depicted in Fig. \ref{fig:bs_metric}, whereas the corresponding scalar and vector field distributions are provided in Fig. \ref{fig:fields}. 

\begin{table}[t!]
\begin{tabular}{|c |c| c| c| c| c|}
\hline
Type &$\gamma_0$ & $\mu M$ & $\mu R$ & $R/M$ & $\omega /\mu$   \\ \hline
BS1 & $0.40$ & $0.609$ & $5.46$ & $8.97$ & $0.811$   \\
BS2 & $0.25$ & $0.632$ & $7.46$ & $11.8$ & $0.864$   \\
BS3 & $0.18$ & $0.612$ & $9.16$ & $15.0$ & $0.896$   \\
BS4 & $0.12$ & $0.572$ & $11.1$ & $19.4$ & $0.922$ \\
\hline
\end{tabular}
\caption{Parameters describing the boson star configurations  \{BS1,BS2,BS3,BS4\}, where $\gamma_0$ denotes the central density of the normalized scalar field radial wave function.}
\label{tab:bs_details}
\end{table}
\begin{table}[t!]
\begin{tabular}{|c|c| c| c| c| c|}
\hline
Type &$f_0$ & $\mu M$ & $\mu R$ & $R/M$ & $\omega/\mu$ \\ \hline
PS1 &$0.210$ & $1.04$ & $9.35$ & $8.99$ & $0.850$ \\
PS2 &$0.092$ & $1.05$ & $12.7$ & $12.1$ & $0.892$ \\
PS3 &$0.057$ & $1.00$ & $15.1$ & $15.0$ & $0.914$ \\
PS4 &$0.033$ & $0.925$ & $18.4$ & $19.9$ & $0.936$ \\
\hline
\end{tabular}
\caption{Parameters describing the Proca star configurations \{PS1,PS2,PS3,PS4\}, where $f_0$ denotes the central value of the time-component of the vector field radial wave function.}
\label{tab:ps_details}
\end{table} 

\begin{table*}[t!]
\begin{tabular}{|c|c |c |c |c |c |c |c |c|}
\hline
Boson star & $\gamma_0$ & $a_1$ & $a_2$ & $a_3$ & $a_4$ & $a_5$ & $a_6$ & $a_7$ \\ \hline 
BS1 & $0.40$ & $-8.38$ & $-1.77$ & $6.08$ & $-0.204$ & $1.32$ & $0.0750$ & $0.0536$ \\
BS2 & $0.25$ & $-5.63$ & $-0.797$ & $6.59$ & $-0.295$ & $0.647$ & $0.0244$ & $0.0412$ \\
BS3 & $0.18$ & $-4.55$ & $-0.457$ & $6.22$ & $-0.163$ & $0.387$ & $0.00991$ & $0.0291$ \\
BS4 & $0.12$ & $-3.78$ & $-0.261$ & $5.63$ & $-0.0354$ & $0.231$ & $0.00378$ & $0.0192$
\\
\hline
\end{tabular}
\ \\
\begin{tabular}{|c|c |c |c |c |c |c |c |c|}
\hline
Boson star & $\gamma_0$ & $b_1$ & $b_2$ & $b_3$ & $b_4$ & $b_5$ & $b_6$ & $b_7$ \\ \hline
BS1 & $0.40$ & $0.269$ & $0.211$ & $0.304$ & $0.290$ & $0.0250$ & $0.209$ & $1.19$ \\
BS2 & $0.25$ & $0.107$ & $0.0492$ & $0.702$ & $0.115$ & $0.0729$ & $0.0346$ & $0.916$ \\
BS3 & $0.18$ & $6.05$ & $1.77$ & $-0.0429$ & $2.15$ & $-0.209$ & $0.764$ & $0.494$ \\
BS4 & $0.12$ & $3.44$ & $0.844$ & $-0.0264$ & $1.32$ & $-0.0993$ & $0.280$ & $0.355$ \\ 
\hline
\end{tabular}
\caption{Parameters $a_i$ and $b_i$ in Eqs. \eqref{sol_grr} and \eqref{sol_gtt} for the scalar boson star configurations detailed in Table \ref{tab:bs_details}.}
\label{tab:bs_pars}
\end{table*}

\begin{table*}[t!]
\begin{tabular}{|c|c| c| c| c| c| c| c| c|}
\hline
Proca star & $f_0$ & $a_1$ & $a_2$ & $a_3$ & $a_4$ & $a_5$ & $a_6$ & $a_7$ \\ \hline 
PS1 & $0.210$ & $-2.71$ & $-1.40$ & $144$ & $-25.7$ & $2.43$ & $0.413$ & $0.644$ \\
PS2 &$0.092$ & $-1.09$ & $-0.999$ & $77.7$ & $-10.9$ & $0.935$ & $0.0785$ & $0.176$ \\
PS3 &$0.057$ & $-1.03$ & $0.384$ & $67.9$ & $-0.660$ & $-0.958$ & $0.138$ & $-0.677$ \\
PS4  & $0.033$ & $-1.48$ & $1.00$ & $422$ & $-35.3$ & $-0.103$ & $0.199$ & $-0.378$\\
\hline
\end{tabular}
\ \\
\begin{tabular}{|c|c| c| c| c| c| c| c| c|}
\hline
Proca star &$f_0$ & $b_1$ & $b_2$ & $b_3$ & $b_4$ & $b_5$ & $b_6$ & $b_7$ \\ \hline
PS1 &$0.210$ & $5.75$ & $1.10$ & $-0.0429$ & $2.15$ & $-0.0625$ & $0.484$ & $0.827$ \\
PS2 &$0.092$ & $23.3$ & $3.06$ & $-0.144$ & $7.19$ & $-0.325$ & $0.827$ & $0.498$ \\
PS3 &$0.057$ & $15.5$ & $1.87$ & $-0.104$ & $5.23$ & $-0.428$ & $0.391$ &$0.368$ \\
PS4 &$0.033$ & $-0.0103$ & $0.00461$ & $0.812$ & $0.00440$ & $0.00281$ & $0.000799$ & $0.361$ \\ 
\hline
\end{tabular}
\caption{Parameters $a_i$ and $b_i$ in Eqs. \eqref{sol_grr} and \eqref{sol_gtt} for the Proca star configurations detailed in Table \ref{tab:ps_details}.}
\label{tab:ps_pars}
\end{table*}

The bosonic star configurations depicted in Tables \ref{tab:bs_details} and \ref{tab:ps_details} are well-described by analytical approximations of their metric components of the form
\begin{equation}\label{sol_gtt}
g_{tt}=-\exp\left\{b_7\left[\exp\left(-\frac{1+b_1x+b_2x^2}{b_3+b_4x+b_5x^2+b_6x^3}\right)-1\right]\right\},
\end{equation}
and
\begin{equation}\label{sol_grr}
g_{rr}=\exp\left\{a_7\left[\exp\left(-\frac{1+a_1x+a_2x^2}{a_3+a_4x+a_5x^2+a_6x^3}\right)-1\right]\right\},
\end{equation}
where $x=\mu r$ is the normalized radial coordinate and the parameters $b_i$ and $a_i$ ($i=1, \ldots, 6$) are constants to be adjusted depending on the values of the boundary conditions $\gamma_0$ and $f_0$ for the scalar boson stars and Proca stars, respectively. The values of the parameters $a_i$ and $b_i$ for each of the four scalar boson star solutions  are found in Table \ref{tab:bs_pars}, and for the four Proca star solutions in Table  \ref{tab:ps_pars}. With these choices for the parameters, the analytical expressions in Eqs. \eqref{sol_gtt} and \eqref{sol_grr} describe the metric components $g_{tt}$ and $g_{rr}$ of each of the eight solutions with errors always smaller than $1\%$ and average relative errors of the order of $0.1\%$ in the interval $0<x<50$. For more information regarding the relative errors, we refer the reader to Appendix \ref{app:errors}. The choice of the boundary $x=50$  corresponds to the numerical infinity considered in the shooting methods, but the upper boundary provided for average relative fitting error is conservative enough to remain valid even if one restricts the analysis to the region where the fundamental fields are localized. Furthermore, given the order of magnitude of the fitting errors, we do not expect these to alter the results in any noticeable qualitative way. This info will feed our analysis of shadows of scalar boson and Proca stars, which we tackle in the next section.

\section{Shadows of boson and Proca stars}\label{sec:shadows}


\subsection{Null geodesic equations and effective potential}

The motion of photons in a gravitational field is given by $g_{ab}\dot{x}^{a}\dot{x}^{b}=0$, where in our case $x^{a}=(t,r,\theta,\phi)$ and dots represent derivatives with respect to the affine parameter $\lambda$ along the geodesics. The static character and the spherical symmetry of the system allows one to conveniently restrict the analysis of the motion to the equatorial plane, $\theta=\pi/2$, without loss of generality,  and defines two conserved quantities: the energy per unit mass, $E=-g_{tt}\dot{t}$, and the angular momentum per unit mass, $L=r^2 \dot{\phi}$. Consequently, the geodesic equation can be written in the familiar form of an equation for the one-dimensional motion of a single particle (re-scaling the affine parameter as $\lambda \to \lambda/L$):
\begin{equation} \label{eq:geoeq}
\dot{r}=V(r),
\end{equation}
where the effective potential reads
\begin{equation} \label{eq:Veff}
V(r)=-g_{tt}^{-1}g_{rr}^{-1} \left(\frac{1}{b^2}+\frac{g_{tt}}{r^2} \right),
\end{equation}
where we have defined the impact parameter as $b \equiv L/E$. A turning point happens when $V(r_0)=0$ for a given $r_0$, i.e., a photon approaching from asymptotic infinity reaches a minimum distance $r_0$ from the compact object before being scattered away back to asymptotic infinity. Moreover, if such a point is located at the minimum of the potential, $V'(r_m)=0, V''(r_m)>0$, then the corresponding curve is dubbed as the photon sphere, which corresponds to the locus of light rays that asymptotically approach a bound unstable orbit.  When this happens, the  associated critical impact parameter, $b_c=\frac{r_m}{\sqrt{-g_{tt}(r_m)}}$, splits the space of rays emitted from a light source into two regions: those with $b>b_c$ are scattered at $r_0$ (and can potentially reach the observer), while those with $b<b_c$ will spiral down until hitting the center of the object (or the event horizon in the case of a black hole). Those lying at $b \gtrsim b_c$ can turn several times around the compact object, producing a characteristic photon ring in the asymptote to $b=b_c$, potentially nesting within it an infinite sequence of stacked light rings depending on both the optical and geometrical assumptions on the accretion flow \cite{Gralla:2019xty}. Note that for a Schwarzschild black hole, $R=2M$, $r_m=3M$, and $b_c=3\sqrt{3}M$. 

The scalar boson  \{BS1,BS2,BS3,BS4\} and Proca  \{PS1,PS2,PS3,PS4\} star configurations considered in the previous section are not compact enough to hold a photon sphere. This is inferred from the behavior of the metric functions $g_{tt}$ and $g_{rr}$ depicted in Figs. \ref{fig:bs_metric}. When plugged in the expression of the effective potential in Eq.(\ref{eq:Veff}) one finds that no minimum is attained. Therefore, on the grounds that boson and Proca stars considered in this work are not capable of producing strong gravitational light deflection to make light rays to wind around them more than one(-half) times (see however \cite{Cunha:2015yba}), in the optical appearance of the object one should expect a single luminous ring made up of those light rays with deflection angles smaller than $\pi/2$, and enclosing a central brightness depression (the shadow). 

\subsection{Disk's emission properties}

In order to verify these expectations we consider an infinitesimally thin accretion disk which emits monochromatically and isotropically with zero absorption \cite{Gold:2020iql}. Hence, it is described by a single luminosity function given by (in the frame of the emitter) $I^e_{\nu_e}=I(r)$, where $\nu_e$ is the frequency of the emitted radiation\footnote{This is admittedly yet another simplification of our analysis in that we are considering a single emission frequency rather than a single detection one, as the latest EHT observations do at the $1.3$mm wavelength \cite{EHT2022PaperI}. To facilitate comparison with those observations, one should filter out the set of observed frequencies $\nu_0(b)$ within the detector's range, an aspect to be implemented in future upgrades of our work}. For the sake of this work, we shall assume the effective source of emission to be located from the inner edge of the disk (peaking its maximum value there) outwards, and furthermore place such an edge at three relevant surfaces  given by those of the Schwarzschild black hole to facilitate the comparison of images with the expectations of GR black holes\footnote{In doing so we are implicitly neglecting any contributions of the backreaction of the (scalar and vector) fields making up the star with the disk's material, since the latter penetrates the inner region to the former. Incorporating such a feature goes well beyond the simplified analytical modelling considered here. More information on the effects of heavy accretion disks can be found in Ref. \cite{Cunha:2019hzj}.}. Such surfaces are the Innermost Stable Circular Orbit (ISCO) for time-like observers, $r_{isco}=6M$, the photon sphere, $r_{ps}=3M$, and the event horizon, $r_{eh}=2M$. Specifically, we consider two models for the ISCO, and one for each of the other two surfaces. Arguably such truncated models for the emission of the disk lack a solid physical justification (particularly given the fact that boson stars do not feature event horizons), but act instead as a proxy for quantifying the differences between black hole and bosonic star images. On the other hand,  it is clear that if the accretion disk is truncated at some finite radius, there will be a dark region in the center of the observer's screen caused not by the properties of the background space-time but by the disk model itself. The truncated disk models should therefore only be taken into consideration to compare the lensing in the outer regions of the observer's screen, and not to draw conclusions about the properties of the shadow itself. In the latest part of our analysis, and to correct this issue, we consider an extra disk model where the disk extends all the way down to the center of the star, $r=0$. Note that, besides the location of the finite radius of the disk at which the emission is truncated, these models are customized in order to have different decays with the radial distance, so as to investigate how the luminosity provided by the disk is distributed around the star.

\begin{figure}[t!]
\includegraphics[width=8.8cm,height=6.0cm]{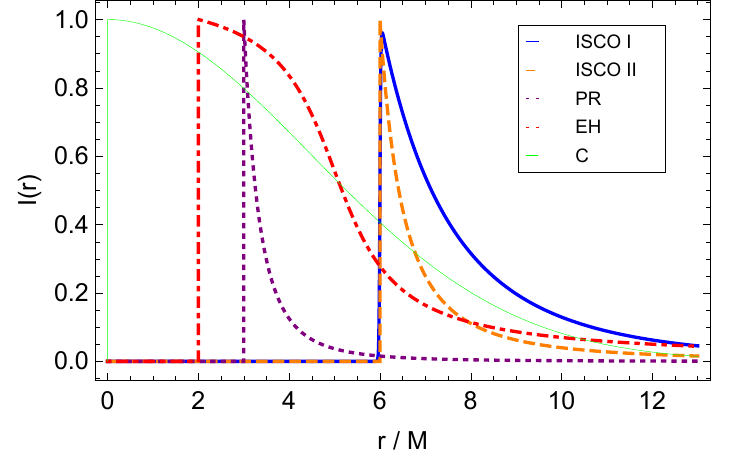}
\caption{The intensity in the emitter's frame, $I(r)$, as a function of the normalized radius, $r/M$, for the five emission profiles in Eqs.(\ref{eq:I1}) to (\ref{eq:I5}). These profiles are labeled by the radius at which the emission starts, taken by convenience to coincide with their Schwarzschild values: the ISCO radius $r/M=6$ (two models), the photon ring, $r/M=3$, and the event horizon, $r/M=2$, alongside with a complementary model for which emission starts at the center of the star, $r=0$, and are customized to have different decays with $r/M$.}
\label{fig:intensity}
\end{figure}

Specifically, these four plus one models are given (in units of $M$), respectively, by
\begin{eqnarray}
I_{isco1}(r)&=& \left(\frac{6}{r}\right)^4 \frac{1+\tanh[50(r-6)]}{2} \label{eq:I1}  \\
I_{isco2}(r)&=&\left\lbrace\begin{array}{l}\frac{1}{(r-5)^2} \hspace{0.1cm} \text{if} \hspace{0.1cm} r \geq 6 \\ 0 \hspace{0.9cm} \text{if} \hspace{0.1cm} r < 6 \end{array}\right. \label{eq:I2} \\
I_{pr}(r)&=&\left\lbrace\begin{array}{l}\frac{1}{(r-2)^3} \hspace{0.1cm} \text{if} \hspace{0.1cm} r \geq 3 \\ 0 \hspace{0.9cm} \text{if} \hspace{0.1cm} r < 3 \end{array}\right.  \label{eq:I3} \\
I_{eh}(r)&=&\left\lbrace\begin{array}{l}\frac{\pi/2-\tan(r-5)}{\pi/2-\tan(-3)} \hspace{0.1cm} \text{if} \hspace{0.1cm} r \geq 2 \\ 0 \hspace{1.8cm} \text{if} \hspace{0.1cm} r < 2 \end{array}\right.  \label{eq:I4} \\
I_{c}(r)&=&e^{-r^2/40} . \label{eq:I5}
\end{eqnarray}
These choices are admittedly somewhat ad hoc, yet they are qualitative similar to those employed by other works in the subject, see e.g. \cite{Gralla:2019xty,Peng:2020wun,Wang:2020emr,Zeng:2020vsj,Li:2021riw}. These five profiles are depicted in Fig. \ref{fig:intensity} where, in addition to the surface at which the emission is started, differences in the slope of its decay are also clear. We point out that both aspects have a critical impact on the image of all the boson stars, as we shall see in what follows. Now, in the frame of the observer the effect of gravitational redshift displaces the emitted frequency to $\nu_o=g_{tt}^{1/2}(r)\nu_e$ and the associated specific intensity to $I_{\nu_o}^o=(\nu_o/\nu_e)^3I_{\nu_e}^e$, which results in a scaling of the total luminosity received at the observer's screen given by \cite{Gralla:2019xty}
\begin{equation} \label{eq:intensity}
I_{o}\equiv \int I^o_{\nu_0}d\nu_0=\int g_{tt}^{3/2}I^e_{\nu_e}d\nu_0=g_{tt}^2(r)I(r)
\end{equation}
which is the main formula for our interest.

To produce the shadow images, we implement a Mathematica-based ray-tracing code to backward integrate the geodesic equation in Eq.(\ref{eq:geoeq}), suitably rewritten as
\begin{equation}
\frac{d\phi}{dr}=\pm \frac{b}{r^2} \frac{\sqrt{-g_{tt}g_{rr}}}{\sqrt{1+\frac{b^2g_{tt}}{r^2}}}
\end{equation}
(where $\mp$ for ingoing and outgoing geodesics, respectively) from the screen of the observer towards the source of emission, and corresponding to a geodesic congruence covering a relevant region of the impact parameter space, $b\in (0,10)$. 

\subsection{Shadows of Schwarzschild black holes}

\begin{figure*}[t!]
\includegraphics[width=7cm,height=5.0cm]{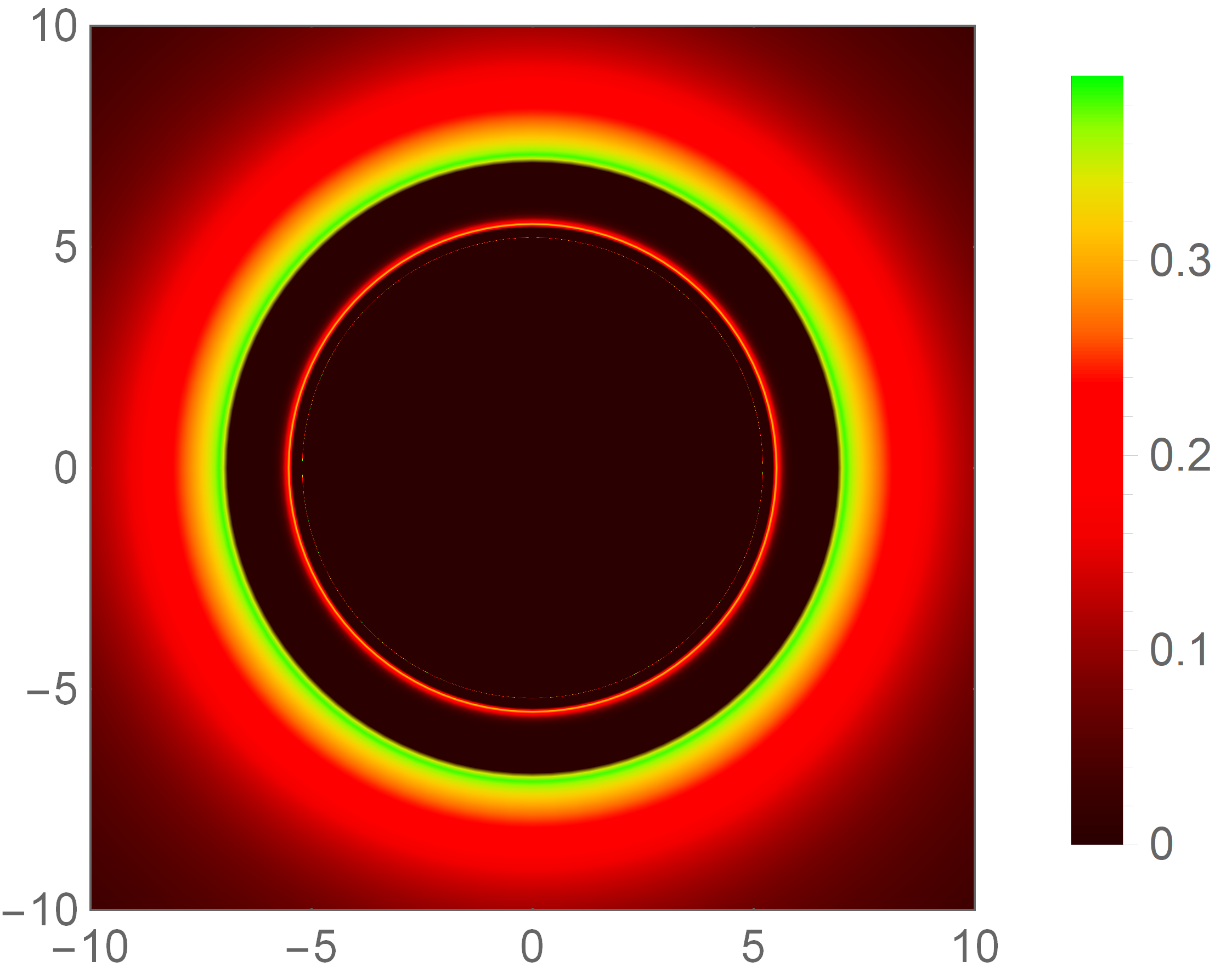}
\includegraphics[width=7cm,height=5.0cm]{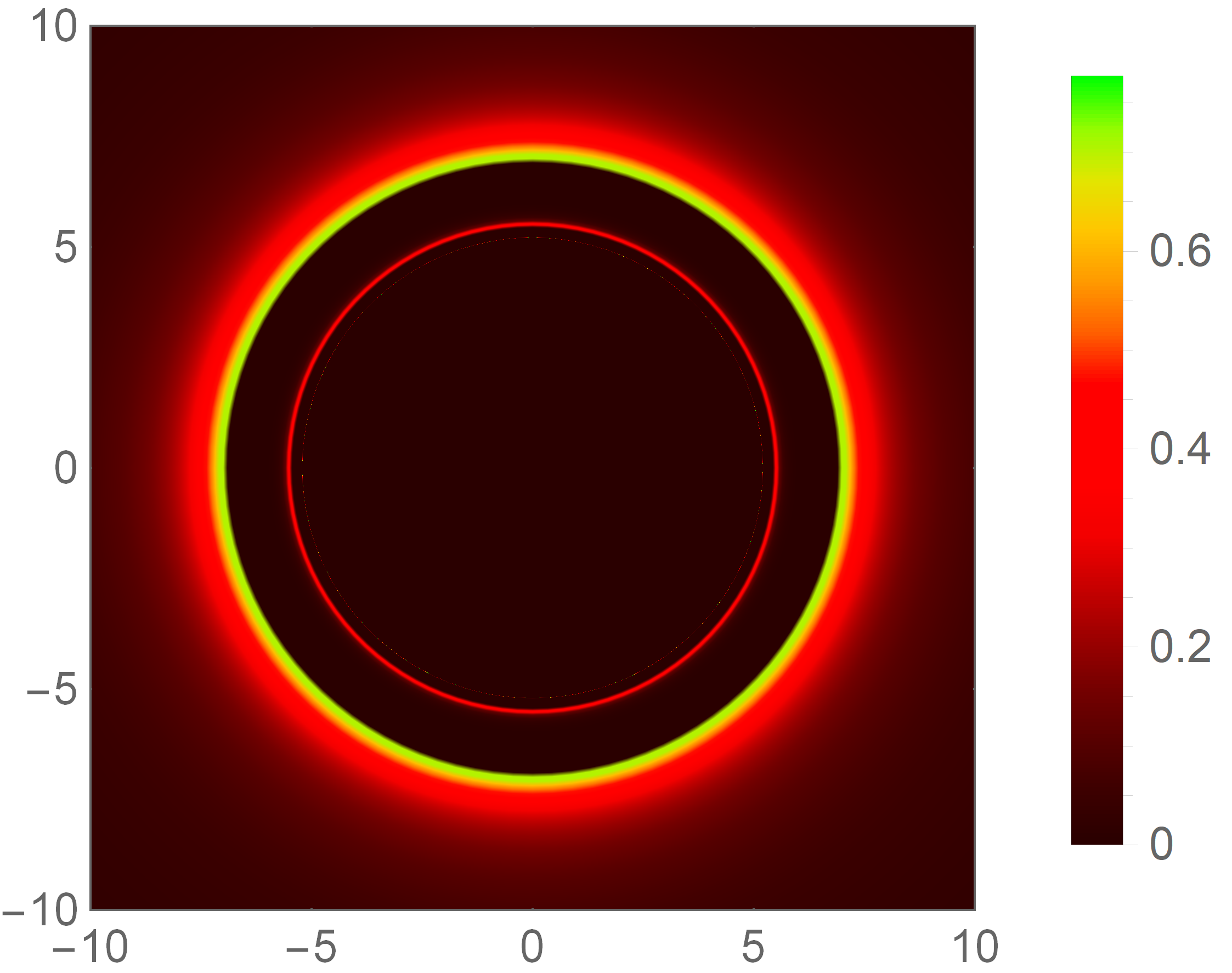}
\includegraphics[width=7cm,height=5.0cm]{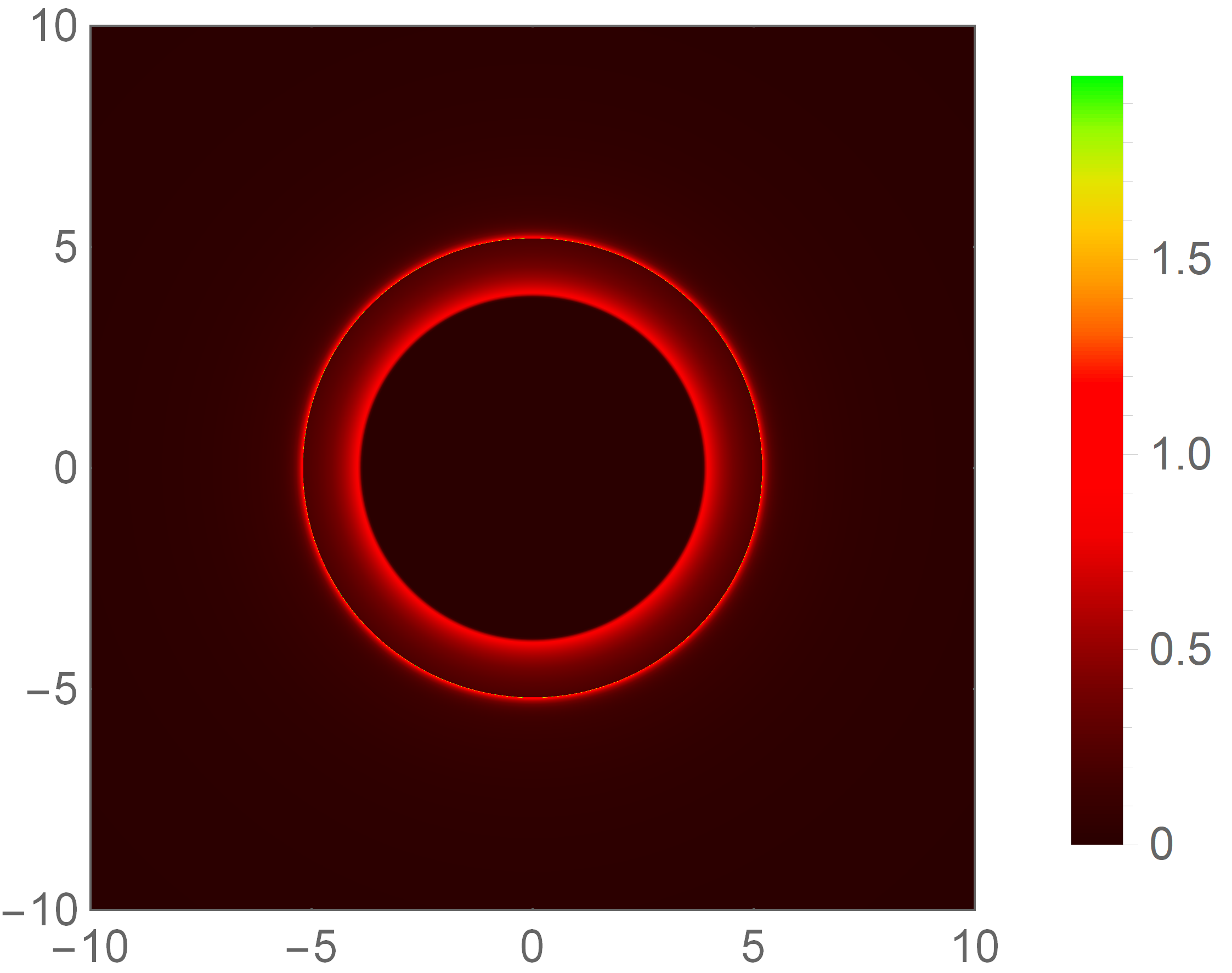}
\includegraphics[width=7cm,height=5.0cm]{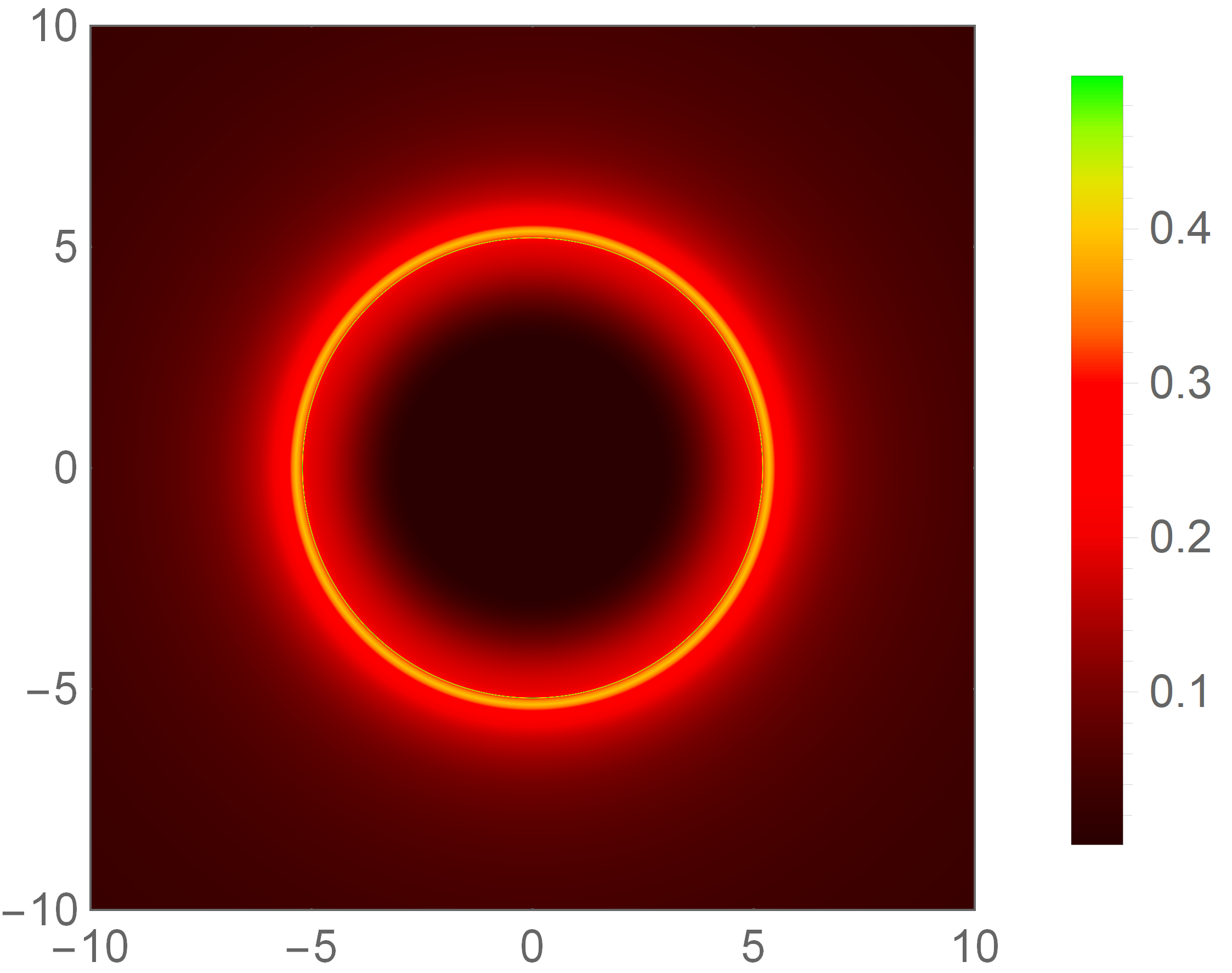}
\caption{The shadows of Schwarzschild black hole using the four accretion models (from left to right and top to bottom) given by Eqs. (\ref{eq:I1}), (\ref{eq:I2}), (\ref{eq:I3}) and  (\ref{eq:I4})   whose effective maximum source of emission is located at the Schwarzschild values for the ISCO, $r=6M$ (first two models), photon sphere, $r=3M$, and at the event horizon radius, $r=2M$. Model (\ref{eq:I5}) does not apply here given the presence of an event horizon.}
\label{fig:SchS}
\end{figure*}

For the sake of comparison with the Schwarschild images, we can recall their main features. In such a case, the fact that a critical curve is present allows light rays to turn several times around the Schwarschild black hole. If the disk is optically thin and its geometry is spherically symmetric, then a thin ring appears on top of the direct emission of the disk, converging to the location of the critical curve itself and bounding the central dark region (the shadow \cite{Falcke:1999pj}). However, if the disk is  geometrically thin or even thick \cite{Vincent:2022fwj}, then this ``photon ring" is instead broken in an infinite sequence of self-similar rings with exponentially-decreasing contributions to the total luminosity of the black hole, in such that the location of the limit of the sequence can penetrate inside the critical curve  provided that the emission profile is also inside it (in our case this would corresponding to the choices (\ref{eq:I4}) and (\ref{eq:I5})). Nonetheless, the presence of a horizon entails that the sequence of such rings cannot converge below a minimum radius dubbed as the  {\it inner shadow} of the black hole \cite{Chael:2021rjo}. The exact details regarding the location and luminosity of such rings depend on the emission properties of the disk which, therefore, conveys a great deal of degeneracy regarding the astrophysical unknowns on the properties of the disk. For the four models of emission given by Eqs. (\ref{eq:I1}), (\ref{eq:I2}), (\ref{eq:I3}) and  (\ref{eq:I4}) [we leave outside model (\ref{eq:I5}) since in this case we have an event horizon] we depict in Fig. \ref{fig:SchS} the corresponding images of Schwarzschild black holes. We see in them that the sub-ring structure and the presence of a (inner) shadow are universal predictions of this setting, though the former can only be seen (at least the first sub-ring beyond the direct emission) in those emission models with a truncated inner edge of the disk far away enough from the horizon (that is, models (\ref{eq:I1}) and (\ref{eq:I2})): otherwise, the sub-ring structure is overlapped with the main direct emission (models (\ref{eq:I3}) and (\ref{eq:I4})) and can hardly be seen.

\begin{figure*}[t!]
\includegraphics[width=4.4cm,height=4.0cm]{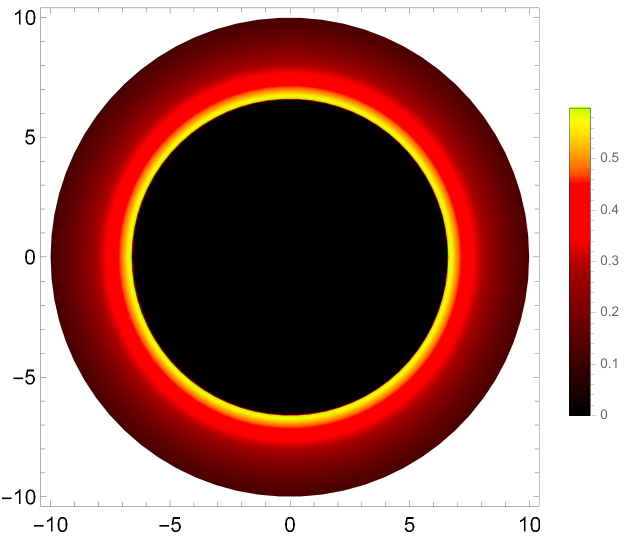}
\includegraphics[width=4.4cm,height=4.0cm]{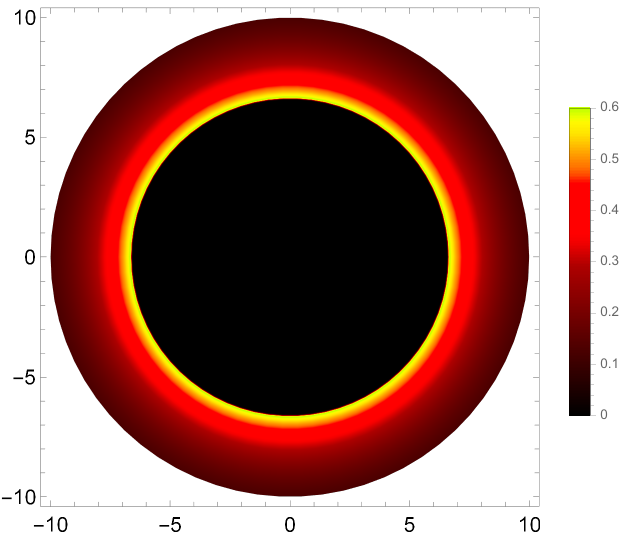}
\includegraphics[width=4.4cm,height=4.0cm]{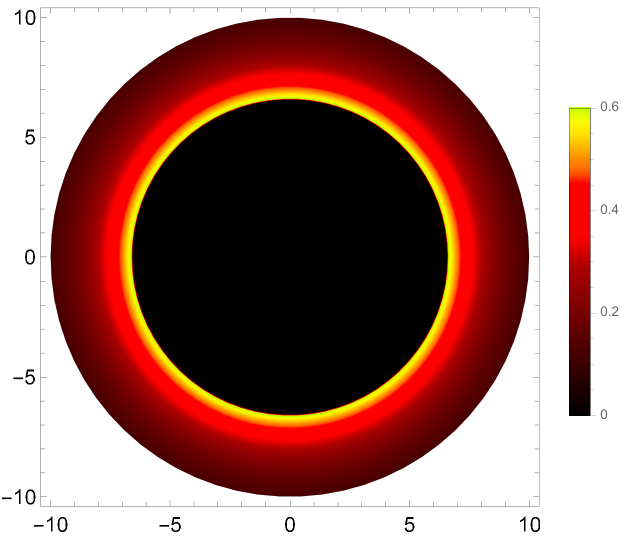}
\includegraphics[width=4.4cm,height=4.0cm]{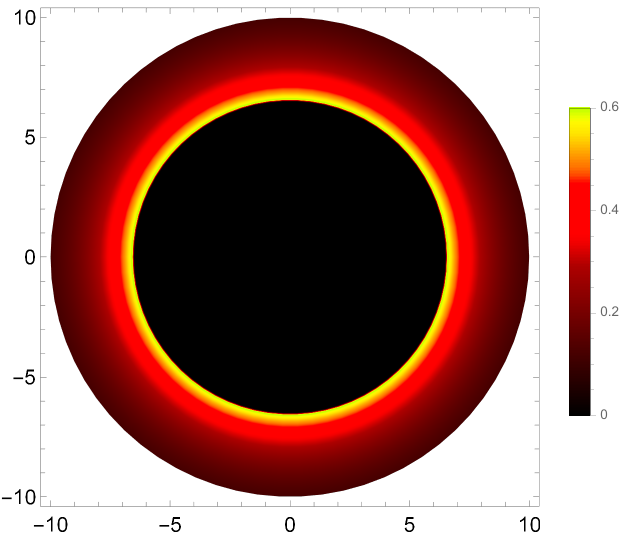}
\includegraphics[width=4.4cm,height=4.0cm]{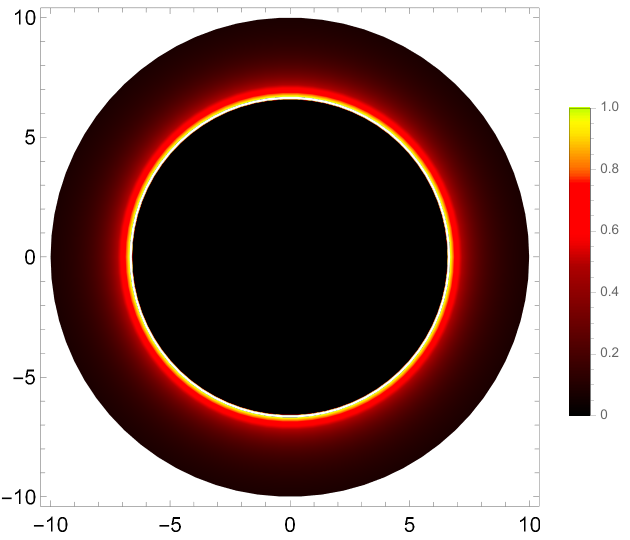}
\includegraphics[width=4.4cm,height=4.0cm]{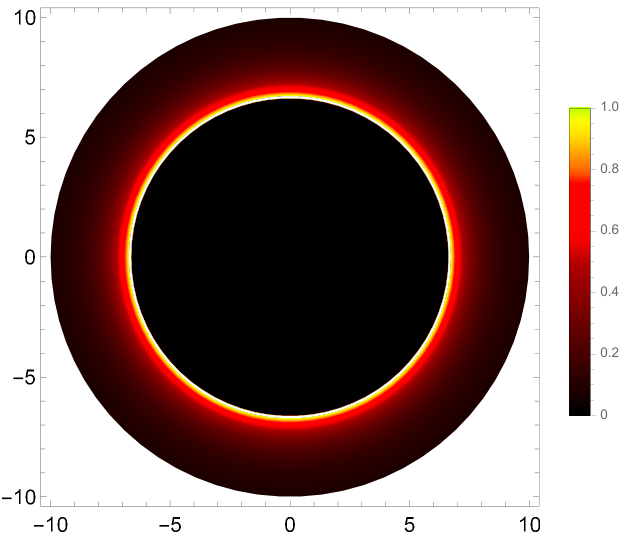}
\includegraphics[width=4.4cm,height=4.0cm]{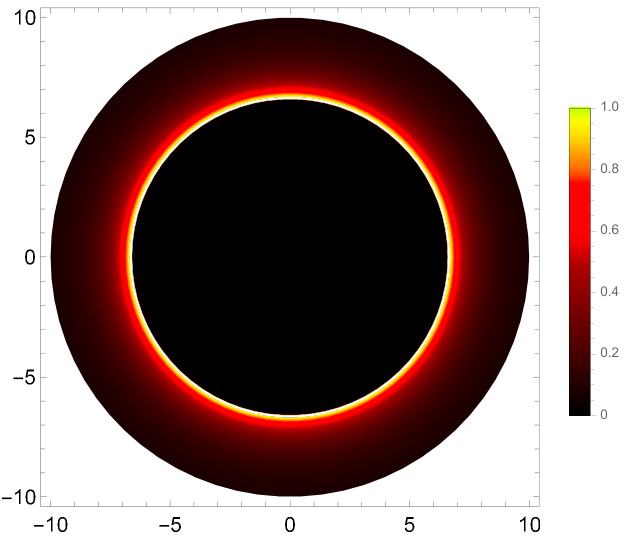}
\includegraphics[width=4.4cm,height=4.0cm]{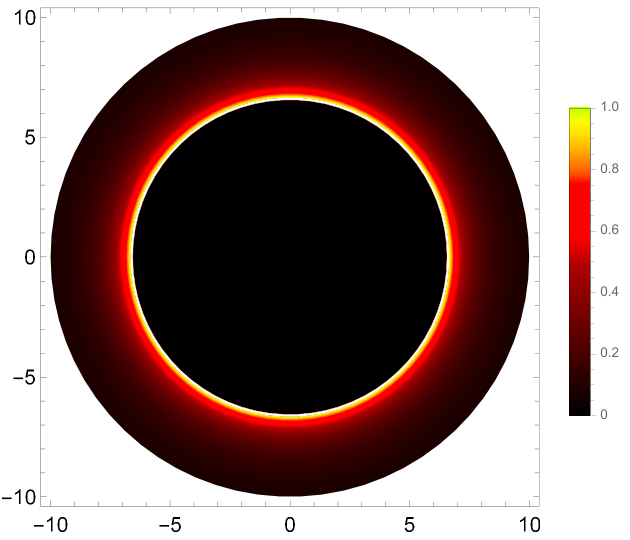}
\includegraphics[width=4.4cm,height=4.0cm]{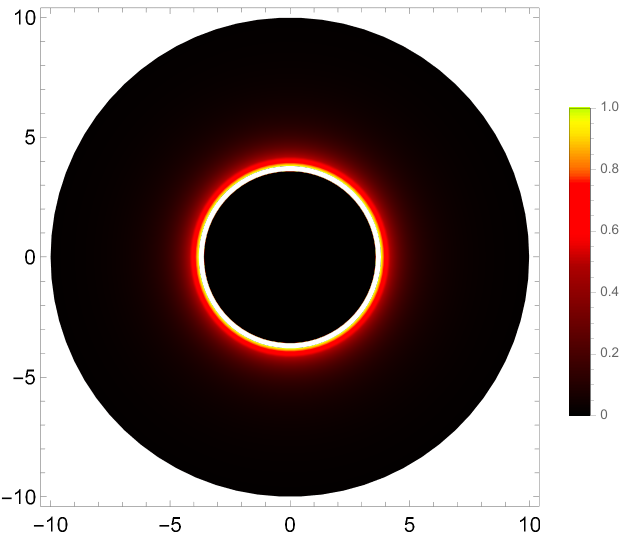}
\includegraphics[width=4.4cm,height=4.0cm]{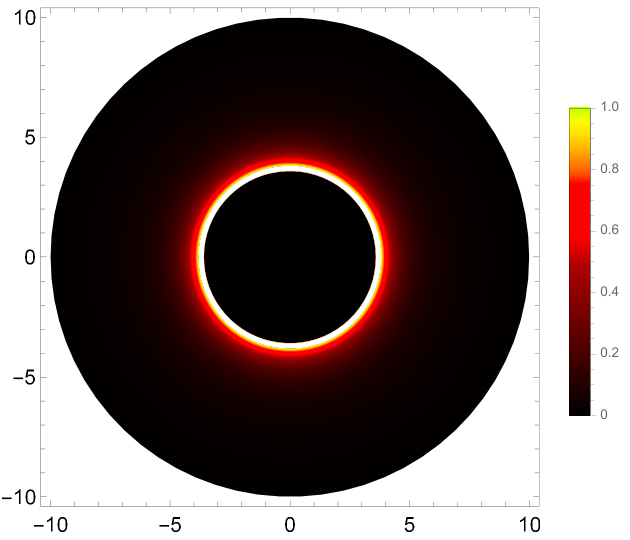}
\includegraphics[width=4.4cm,height=4.0cm]{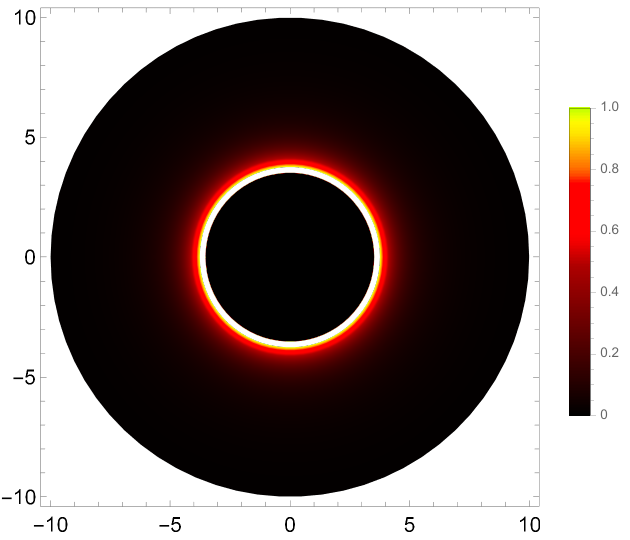}
\includegraphics[width=4.4cm,height=4.0cm]{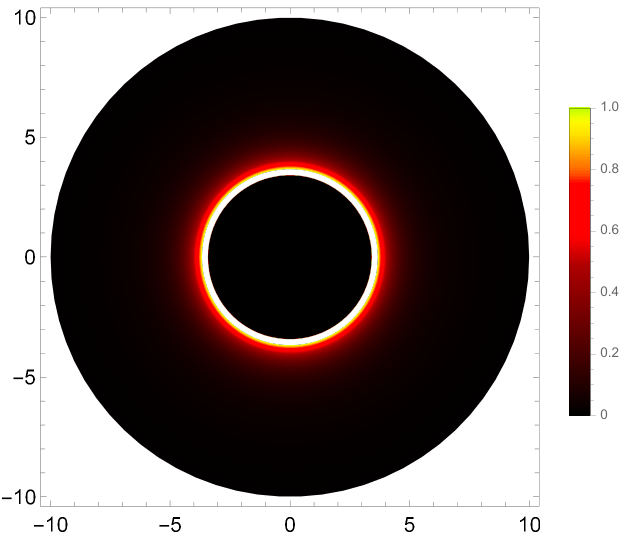}
\includegraphics[width=4.4cm,height=4.0cm]{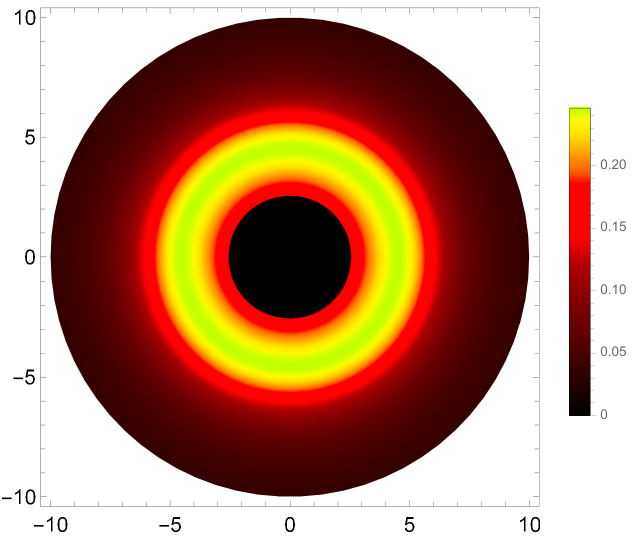}
\includegraphics[width=4.4cm,height=4.0cm]{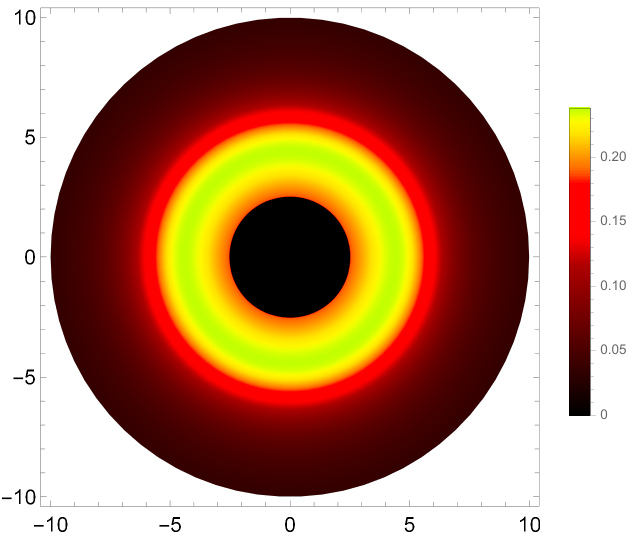}
\includegraphics[width=4.4cm,height=4.0cm]{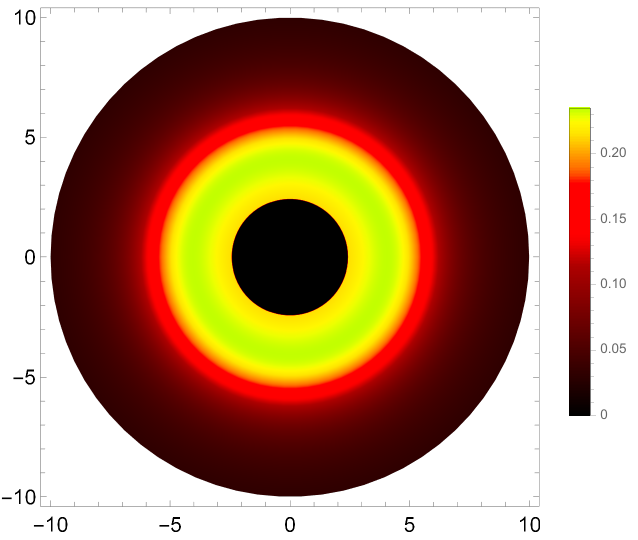}
\includegraphics[width=4.4cm,height=4.0cm]{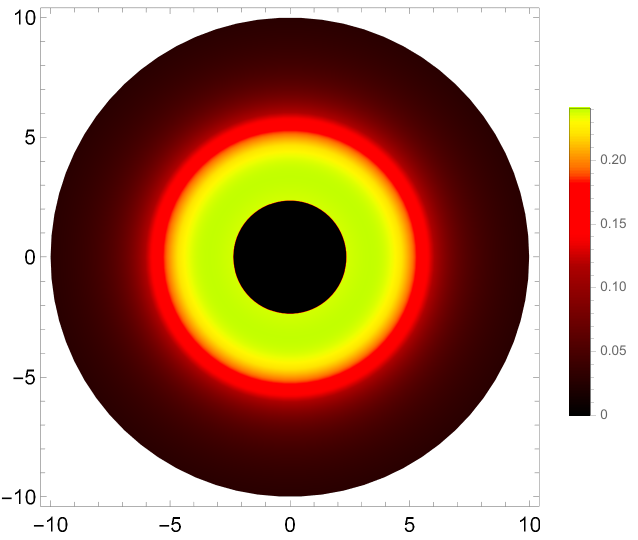}
\includegraphics[width=4.4cm,height=4.0cm]{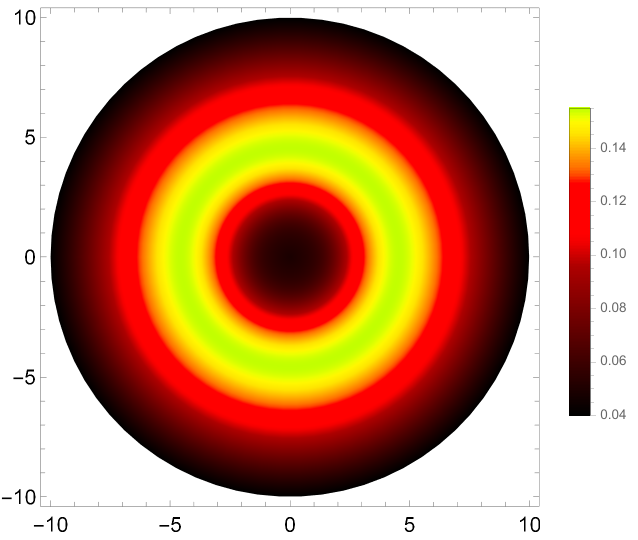}
\includegraphics[width=4.4cm,height=4.0cm]{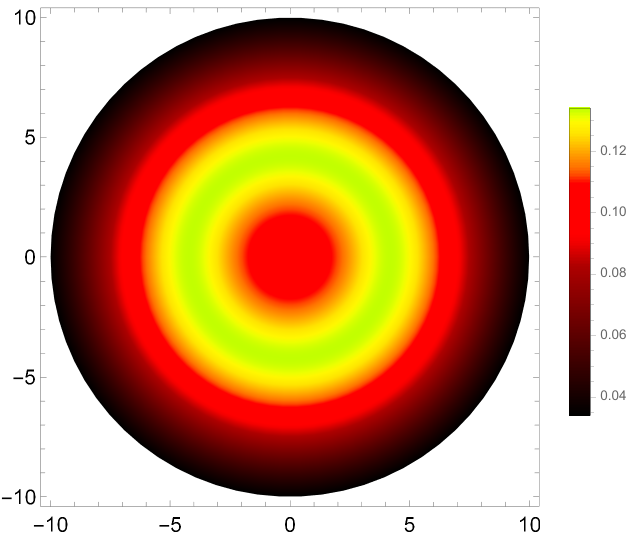}
\includegraphics[width=4.4cm,height=4.0cm]{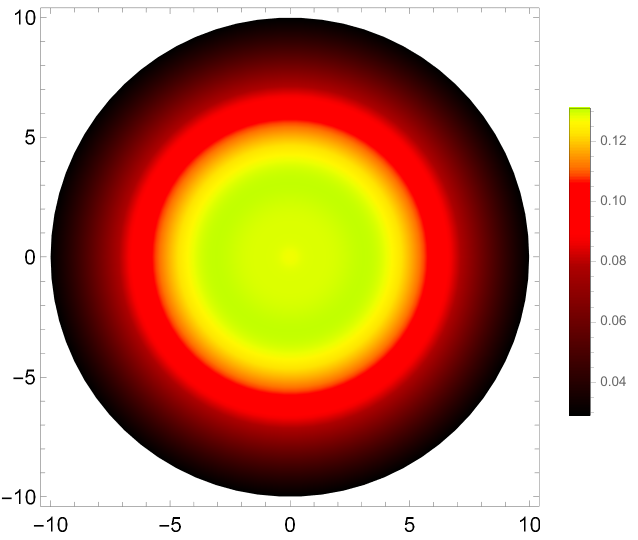}
\includegraphics[width=4.4cm,height=4.0cm]{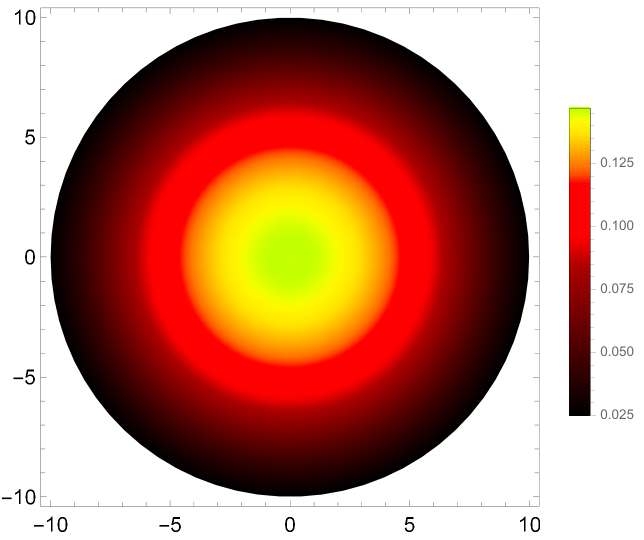}
\caption{The shadows of the boson star configurations (from left to right)  \{BS1,BS2,BS3,BS4\} using the five accretion models (from top to bottom) given by Eqs. (\ref{eq:I1}), (\ref{eq:I2}), (\ref{eq:I3}), (\ref{eq:I4}) and (\ref{eq:I5}),  whose effective maximum source of emission is located at the Schwarzschild values for the ISCO, $r=6M$ (first two models), photon sphere, $r=3M$, event horizon radius, $r=2M$, and at center of the object, $r=0$, respectively.}
\label{fig:shadowBS}
\end{figure*}
\begin{figure*}[t!]
\includegraphics[width=4.4cm,height=4.0cm]{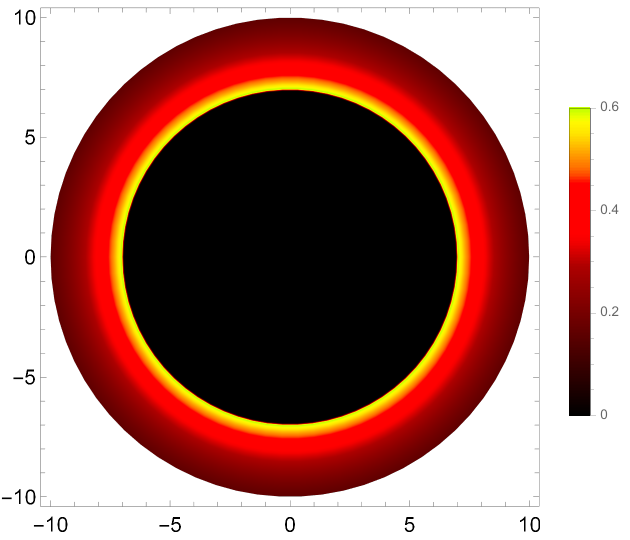}
\includegraphics[width=4.4cm,height=4.0cm]{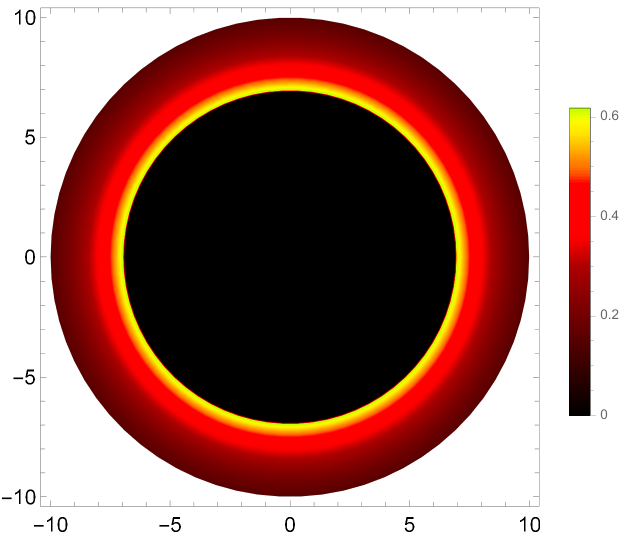}
\includegraphics[width=4.4cm,height=4.0cm]{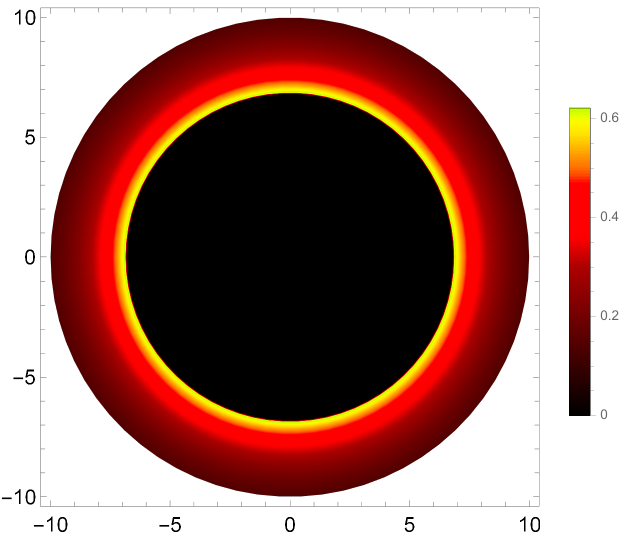}
\includegraphics[width=4.4cm,height=4.0cm]{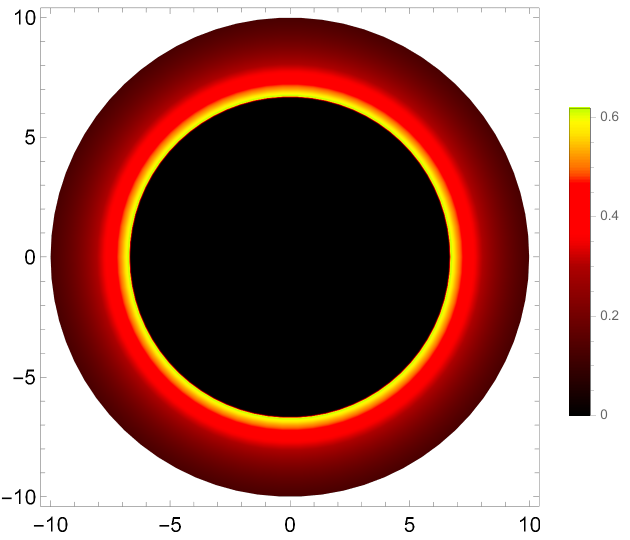}
\includegraphics[width=4.4cm,height=4.0cm]{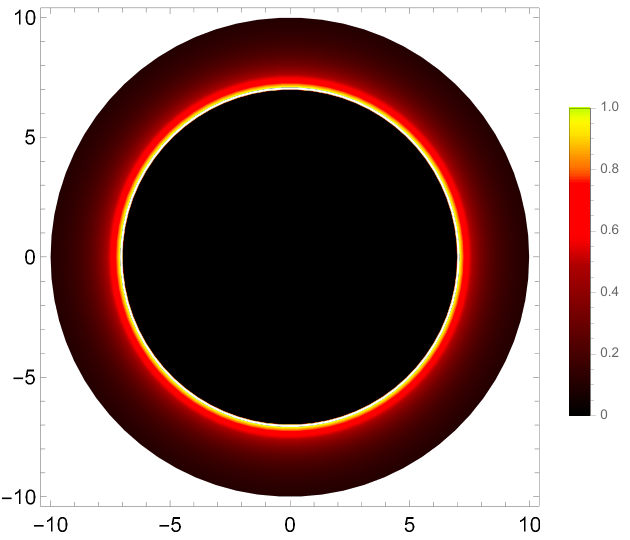}
\includegraphics[width=4.4cm,height=4.0cm]{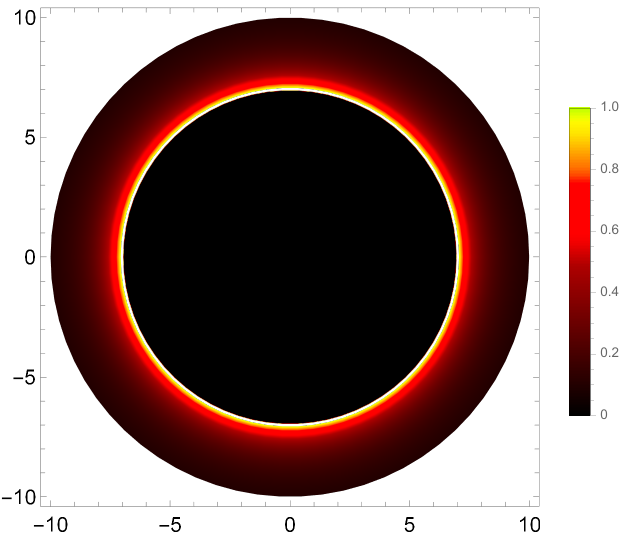}
\includegraphics[width=4.4cm,height=4.0cm]{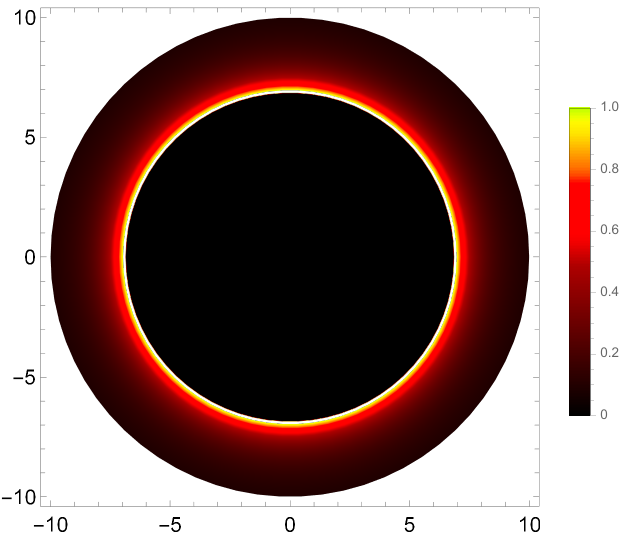}
\includegraphics[width=4.4cm,height=4.0cm]{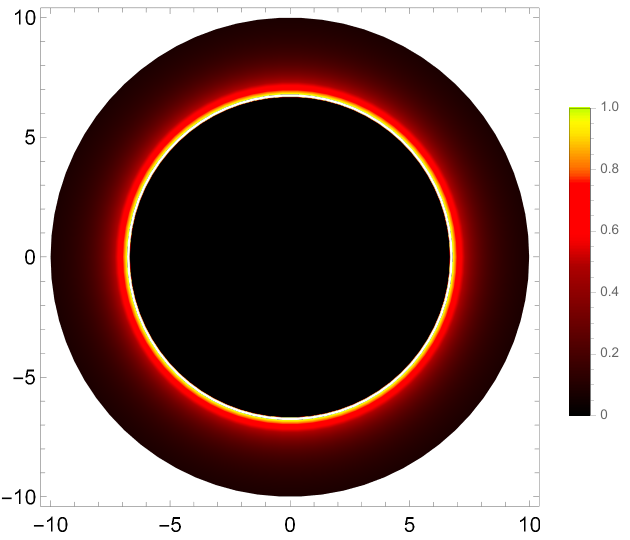}
\includegraphics[width=4.4cm,height=4.0cm]{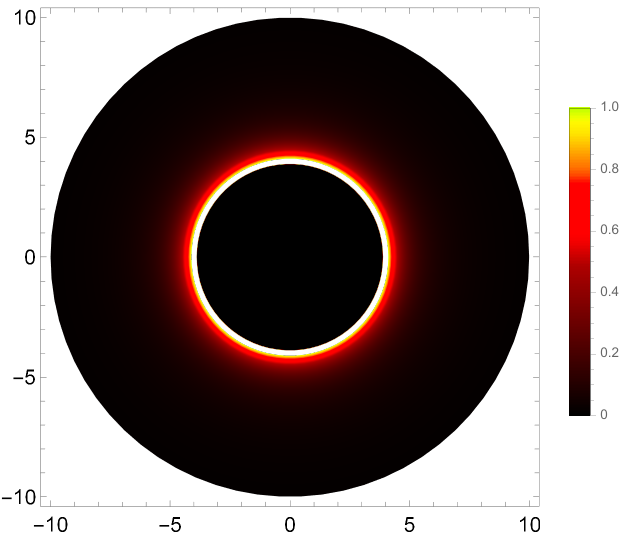}
\includegraphics[width=4.4cm,height=4.0cm]{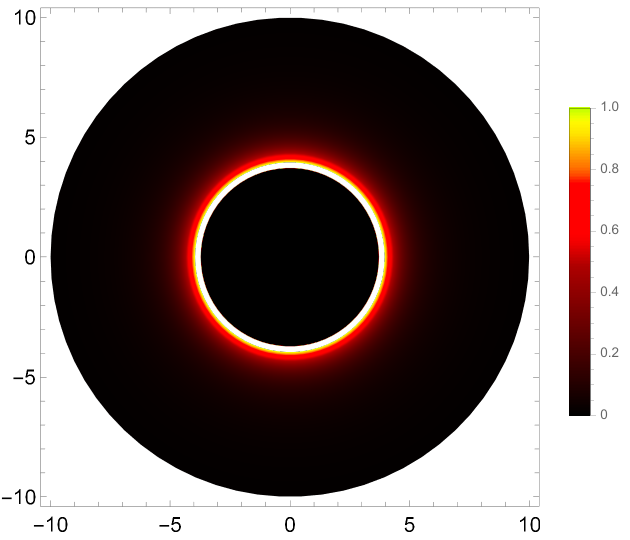}
\includegraphics[width=4.4cm,height=4.0cm]{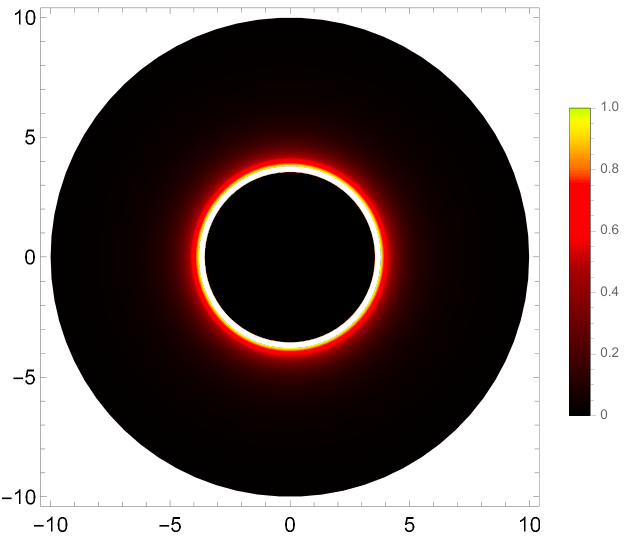}
\includegraphics[width=4.4cm,height=4.0cm]{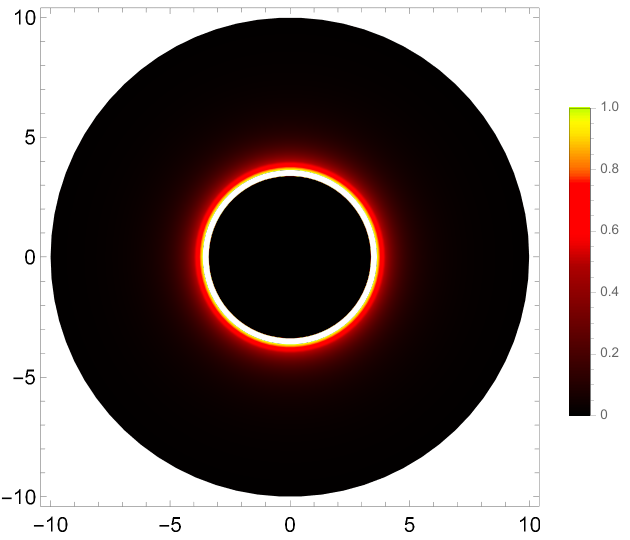}
\includegraphics[width=4.4cm,height=4.0cm]{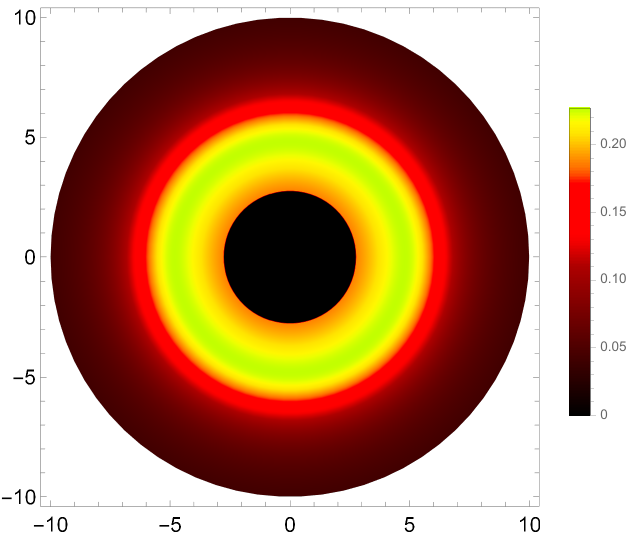}
\includegraphics[width=4.4cm,height=4.0cm]{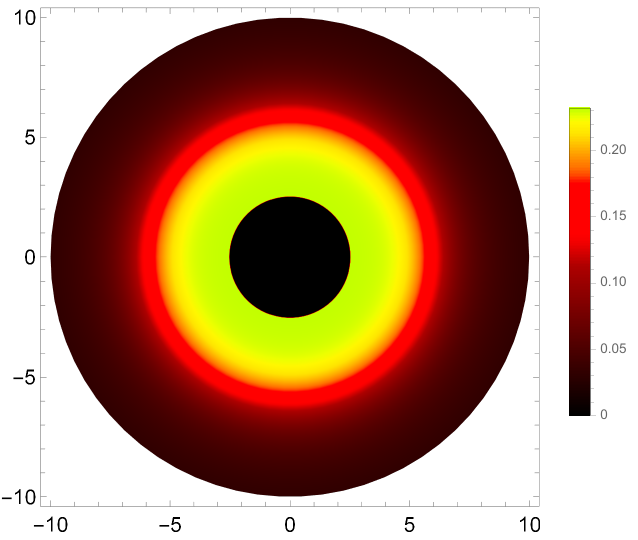}
\includegraphics[width=4.4cm,height=4.0cm]{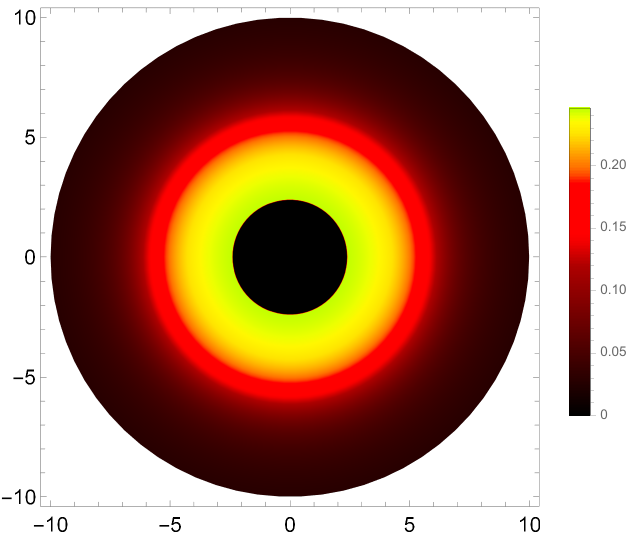}
\includegraphics[width=4.4cm,height=4.0cm]{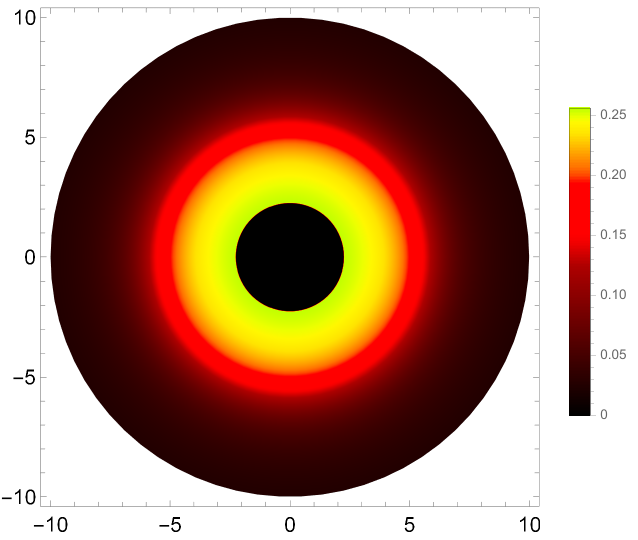}
\includegraphics[width=4.4cm,height=4.0cm]{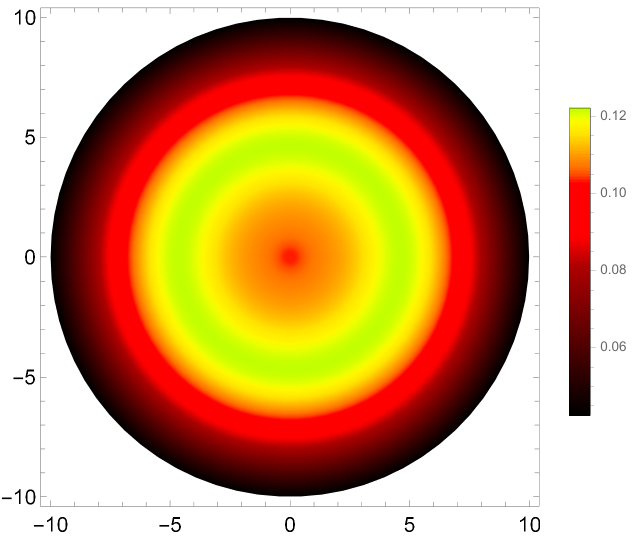}
\includegraphics[width=4.4cm,height=4.0cm]{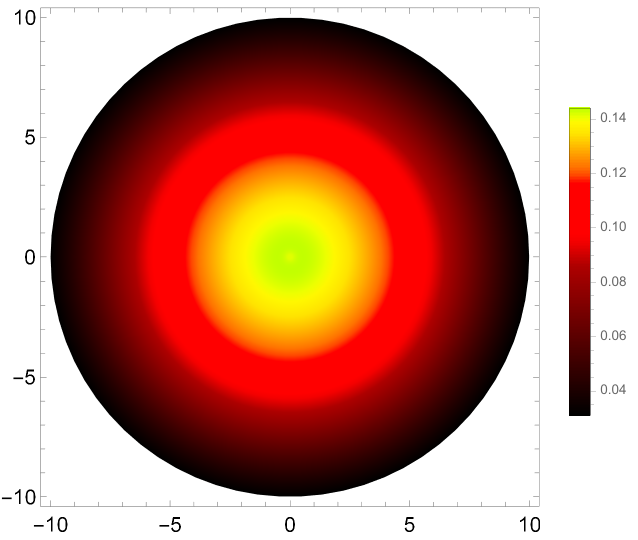}
\includegraphics[width=4.4cm,height=4.0cm]{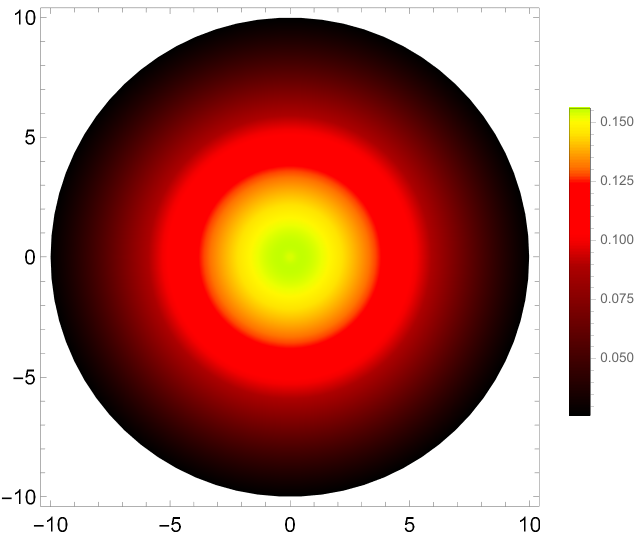}
\includegraphics[width=4.4cm,height=4.0cm]{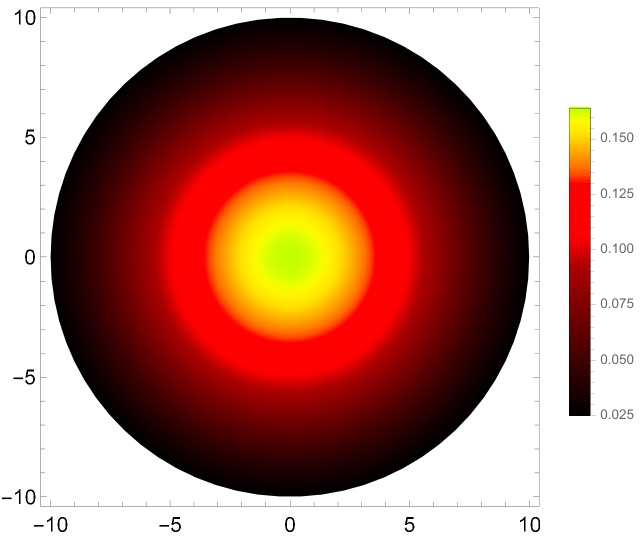}
\caption{The shadows of the Proca star configurations (from left to right)  \{PS1,PS2,PS3,PS4\} using the four accretion models (from top to bottom) given by Eqs. (\ref{eq:I1}), (\ref{eq:I2}), (\ref{eq:I3}), (\ref{eq:I4}) and (\ref{eq:I5}), whose effective maximum source of emission is located at the Schwarzschild values for the ISCO, $r=6M$ (first two models), photon sphere, $r=3M$, event horizon radius, $r=2M$, and at center of the object, $r=0$, respectively }
\label{fig:shadowPS}
\end{figure*}

\subsection{Shadows of boson and Proca stars}

When considering the results of our simulations for boson and Proca stars,  we observe that the gravitational light bending is smaller as we move  through the sequence of solutions \{BS1,BS2,BS3,BS4\}  and  \{PS1,PS2,PS3,PS4\}, respectively, and stronger in the scalar case than in the vector one, which is in agreement with previous observations made in Ref. \cite{Cunha:2015yba}. After the ray-tracing is performed, a shadow-producer code feeds this information into the luminosity given by Eqs.(\ref{eq:I1}) to (\ref{eq:I5}) to {\it colour} scalar boson and Proca stars, bearing in mind the gravitational redshift effect of Eq.(\ref{eq:intensity}). The resultant shadow images for scalar boson stars are provided in Fig. \ref{fig:shadowBS} using the samples \{BS1,BS2,BS3,BS4\}, whereas for Proca stars are given in Fig. \ref{fig:shadowPS} using the samples \{PS1,PS2,PS3,PS4\}, respectively.

Several features are noticeable from these plots. We first verify the existence of a single luminous ring enclosing the central brightness depression for the four emission models truncated at the Schwarzschild relevant surfaces, in agreement with our expectations from the absence of a critical curve in these configurations. This implies that those higher-order rings associated to light trajectories winding several times around the compact object when a photon sphere is present (as discussed above), are absent in shadow images of the scalar boson and Proca stars considered here, which we remark arise as static and spherically symmetric solutions, which are stable against linear perturbations, and without higher-order self-interaction terms. It is important to note, however, these conclusions might change if any of these assumptions is dropped, since in such a case one could find solutions for ultra-compact bosonic stars with light rings, which have recently been shown to be potentially affected by instabilities \cite{Cunha:2022gde}. In any case, for the present solutions without light rings only the direct emission is present, the location of the effective maximum of emission in the luminosity models employed here, after accounting for the gravitational redshift, determines the size of the shadow of these objects. This is somewhat a different situation as in the Schwarzschild black hole, where the presence of the additional emissions associated to higher-order light trajectories (in optically thin scenarios) allow to pierce further into the impact parameter region \cite{Chael:2021rjo}. Therefore, the size of the shadow in scalar boson and Proca stars is comparatively larger than in their same-mass Schwarzschild solution counterpart for the same emission model, which in combination with the slope of its decay it governs the optical appearance of the object in terms of the width and luminosity distribution of the bright ring enclosing it. In this sense, it is worth pointing out that the choice of the metric functions for the bosonic star configurations, within the samples above, has a minor influence in the spacial distribution of the luminosity of the ring. Indeed, the differences for the first two disk models (\ref{eq:I1}) and (\ref{eq:I2})  are only perceivable through a change in the color scales of the figures, while for the third disk model (\ref{eq:I3}) are completely imperceptible. For the fourth model (\ref{eq:I4}), these differences between the observations of different samples of boson/Proca stars become more evident, especially for the most compact configurations. Note also that for the models in Eqs.\eqref{eq:I3} and \eqref{eq:I4} the only qualitative difference between the images obtained for the black hole case and the bosonic star cases is the existence of the ring structure in the former  (recall Fig. \ref{fig:SchS}). If not for this feature, the intensity of the main direct-emission structure is qualitatively and quantitatively similar (see the scales of these figures), suggesting that these configurations would be indistinguishable if the experiments are not precise enough to detect the ring structure. Therefore, given the large uncertainties in the emission properties of accretion disks, it is unclear whether images of Proca/boson stars might be mistaken with those of Schwarzschild black holes in emission models where the inner shadow is not vanishing and the sequence of light rings may be hidden in the direct emission, thus being hard to observe. 

Things change moderately in the fifth emission model (\ref{eq:I5}). In the Schwarzschild black hole images, the minimum size of the central brightness depression that can be achieved corresponds to an accretion disk model in which the emission extends all the way down to the horizon, yielding the inner shadow \cite{Chael:2021rjo}. For the boson scalar and Proca stars considered here, the absence of an event horizon allows for the possibility of extending the inner edge of the disk further, even up to the center of the object itself,  as given by the model in Eq.(\ref{eq:I5}). Interestingly, for the configurations BS1, BS2 and PS1, one verifies that, even though the emission attains a maximum at $r=0$, the observed image features a dimming of luminosity near the center. In particular, for the BS1 configuration, the observed image in this situation resembles qualitatively the image produced by a disk with a peak of emission at $r=2M$. This effect is progressively dimmer as one decreases the compactness of the boson stars, eventually producing images resembling a single luminosity bubble. This result indicates that, even if the emission of the disk is not truncated at some finite radius, horizonless objects without light rings can also produce observable images which are qualitatively similar to those created by the direct emission in conventional black hole shadows if they are compact enough. This result contrasts with Refs. \cite{Olivares:2018abq} and \cite{Herdeiro:2021lwl}, in the sense that in the first of these references the dimming of the luminosity is caused by a decrease in the emission due to a decreasing orbital angular velocity as one approaches the center of the bosonic star, and in the second of these references the accretion disk was truncated at a finite radius, similarly to the first four accretion disk models considered in this work. This way, even if the emission extends all the way down to the center of the bosonic star, a dimming caused by the gravitational redshift would still be observed if the bosonic star is compact enough.

The bottom line of this discussion is that the image of a bosonic star configuration, viewed in face-on orientation, is largely dominated by the emission properties (location of its peak and slope of the decay) of the accretion disk, with the shape of the background geometry playing a secondary role, the exception being the case for which the emission peaks at the center of the object since in such a case the brightness of the central bubble moderately changes from one background geometry to another. In this sense, a combination of the closeness of the inner edge of the disk to the center of the solution combined with a smoother decay of the emission yields a more diluted luminosity in the impact parameter space.  Therefore, the chances to confirm the existence of any such objects relies not only on an accurate understanding of the astrophysics of the former in order to produce images of the latter that can be compared to observations, but also in an upgrade of the current facilities to observe diluted luminosities associated to the potential existence of light rings (and to eventually discard such an existence). 

\section{Conclusion}\label{sec:concl}

In this work we have studied the optical appearance of bosonic stars, i.e., self-gravitating configurations of scalar and vector (Proca) fields. Specifically, we analyzed static and spherically symmetric solutions for bosonic stars without higher-order self interactions and stable against linear perturbations. Within these assumptions, the solutions obtained are not compact enough to hold a photon sphere, that is, an unstable critical orbit to which light rays can asymptotically approach. Consequently, the shadow cast by these objects is quite different from the one of a black hole: while in the latter and depending on the optical properties of the accretion disk one can find a sub-ring structure formed by those light rays that have winded several times around the black hole, nested in the direct emission, and converging to the critical curve (in spherically symmetric scenarios) or penetrating inside it up to the inner shadow (in geometrically thin scenarios), in the former only the direct emission is available to colour the boson star.

We have considered four plus four samples of spherically symmetric scalar and vector boson star configurations, respectively, which correspond  to an analytical fitting for the metric components of a numerical  integration of the field equations,  featuring average relative errors of the order of $\sim 0.1\%$ in the relevant range of the radial coordinate. Next we have considered an infinitely thin accretion disk emitting monochromatically from the finite radius at which the emission is truncated outwards, and considered five different emission profiles with different  locations for this radius where the luminosity takes its maximum value, and different decay slopes with the distance. The results of our simulations for all the combinations between the background geometries and the accretion disk models  are in agreement with the expectations based on the absence of a photon sphere: a single bright ring enclosing the central blackness depression (the shadow) of the object. For the face-on observations considered, the size of this shadow, as well as the luminosity and depth of the bright region, are heavily influenced by both the location of the truncated finite radius and the decay of the emission with the distance, while the choice of the sample for the bosonic star configuration has a sub-dominant influence. Nonetheless, these degenerated observations could become distinguishable for an inclined (as opposed to face-on) observation \cite{Herdeiro:2021lwl}, a topic we are currently addressing and hope to report in the future.

In addition, we have verified that if one models an accretion disk extending close enough to the center of the object such that it features an emission profile peaking at the origin, the background configuration starts to have a non-negligible impact on the produced image. In particular, for the most compact configurations, i.e., BS1, BS2 and PS1, one clearly observes a dimming of the luminosity close to the center of the image, that for the most compact model (BS1) has strong resemblances with a conventional black hole shadow created by its direct emission alone (i.e. without the light rings structure). This effect is progressively diminished by a decrease in the compactness of the configuration considered, eventually leading to images featuring a single bubble of luminosity at the center. This result is of major importance as it shows that horizonless compact objects without a photon sphere are able to produce shadow-like observable images without having to truncate the emission profile at some finite radius, and thus some of these boson stars could be potentially consistent with current and future observations. It is important to note, however, that in our analysis we have neglected the backreaction of the disk's material in the background space-time. Indeed, if matter is being continuously accreted by the bosonic star, one would expect that it will pile up at the center of the star, therefore overruling the backreaction hypothesis and potentially inducing instabilities on the solutions. An upgrade of our work should therefore incorporate this aspect into the very structure of the boson stars background solutions.

To conclude, the EHT observations of a bright unresolved ring of radiation, while being compatible with our rough expectations of a compact object illuminated by its accretion disk,  do not allow to safely conclude the nature of the shadow caster yet, still being potentially compatible with a myriad of possibilities. Nonetheless, with the fast progress of the capabilities of very long baseline interferometry, the chances to search for the intimacy of the ring structure in shadow observations are promising \cite{Johnson:2019ljv}. Whether one would find the expected sub-ring structure foreseen by the Kerr solution, additional light rings associated to more than a single critical curves in black hole/wormhole scenarios, or will just confirm the main ring associated to the direct emission (and nothing else) ascribed to solutions without any critical curve such as the bosonic stars studied in this work, remains to be seen. Therefore, we are at the stage in which we could be able to either further strengthen the position of the Kerr solution as the sole actor of the cosmic ultra-compact cast or to extend the number of viable alternatives for that purpose. To this end, we are currently working on upgrading our shadow images to account for arbitrary inclinations with respect to the observer's light of sight and, at the same time, on verifying the viability of these same bosonic star samples in the context of orbital motion near the galactic center. \\

\begin{acknowledgements}

JLR is supported by the European Regional Development Fund and the programme Mobilitas Pluss (MOBJD647), and thanks the Department of Theoretical Physics at the Complutense University of Madrid for their hospitality during the elaboration of this work. DRG is funded by the \emph{Atracci\'on de Talento Investigador} programme of the Comunidad de Madrid (Spain) No. 2018-T1/TIC-10431, and acknowledges further support from the Ministerio de Ciencia, Innovaci\'on y Universidades (Spain) project No. PID2019-108485GB-I00/AEI/10.13039/501100011033 (``PGC Generaci\'on de Conocimiento") and the FCT projects No. PTDC/FIS-PAR/31938/2017 and PTDC/FIS-OUT/29048/2017.  This work is also supported by the project PROMETEO/2020/079 (Generalitat Valenciana), and the Edital 006/2018 PRONEX (FAPESQ-PB/CNPQ, Brazil, Grant 0015/2019). This article is based upon work from COST Action CA18108, supported by COST (European Cooperation in Science and Technology). All images of this paper were obtained with our own codes implemented within Mathematica\circledR.

\end{acknowledgements}


\appendix

\section{Relative errors of numerical integrations}\label{app:errors}

\begin{figure*}
\includegraphics[scale=0.8]{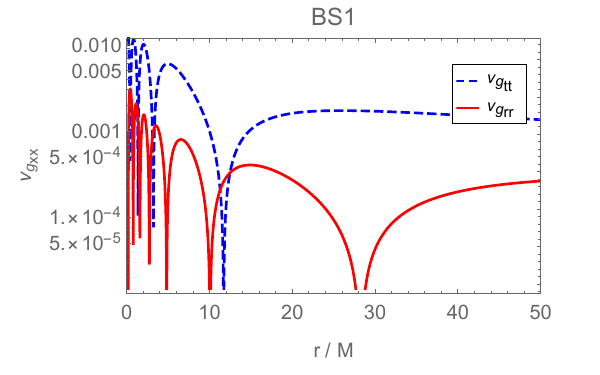}
\includegraphics[scale=0.8]{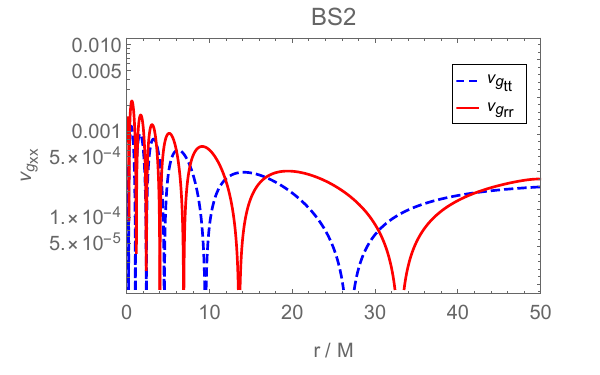}
\includegraphics[scale=0.8]{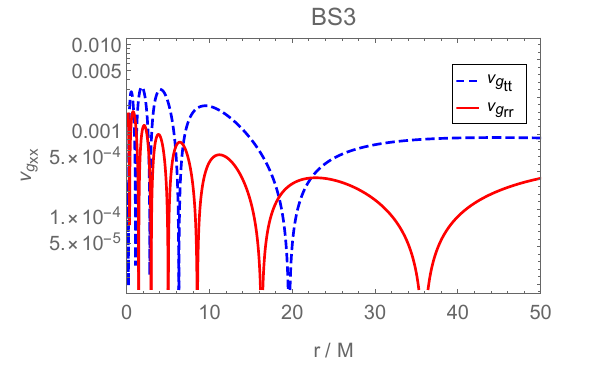}
\includegraphics[scale=0.8]{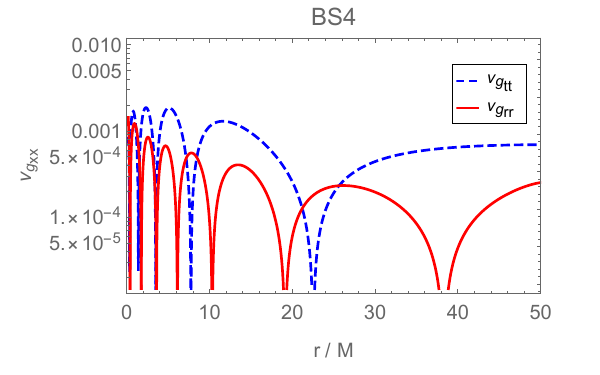}
\includegraphics[scale=0.8]{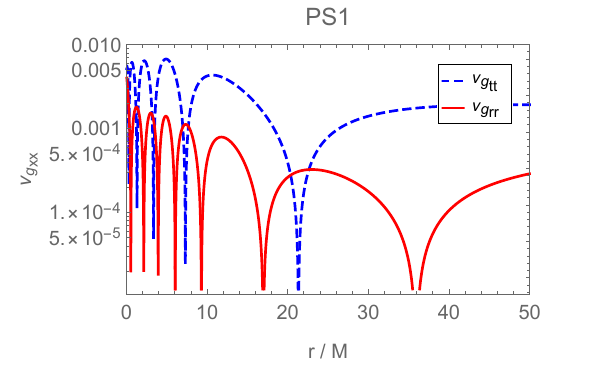}
\includegraphics[scale=0.8]{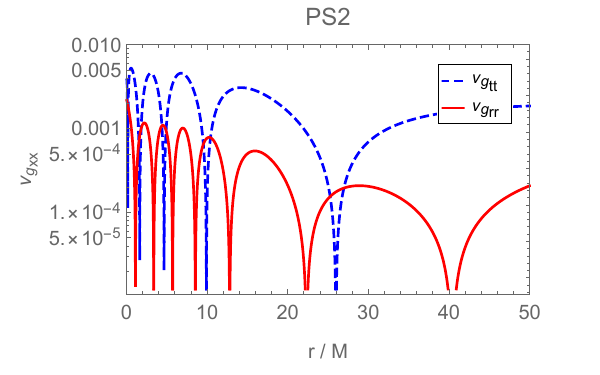}
\includegraphics[scale=0.8]{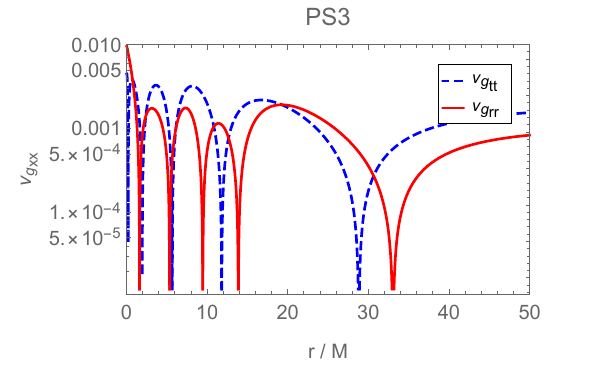}
\includegraphics[scale=0.8]{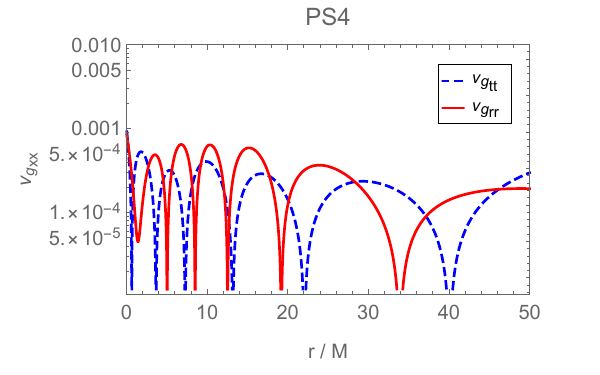}
\caption{Relative errors $\nu_{g_{xx}}$ for the eight bosonic star configurations considered in Sec. \ref{sec:solutions}. All relative errors are always smaller than $1\%$, with average relative errors in the region $0<x<50$ of the order of $0.1\%$.}
\label{fig:errors}
\end{figure*}

In Sec. \ref{sec:solutions} we have introduced eight bosonic star configurations, described by the parameters appearing in Tables \ref{tab:bs_details} and \ref{tab:ps_details}. These numerical solutions were then approximated by the analytical fits for the metric components $g_{tt}$ and $g_{rr}$ given in Eqs. \eqref{sol_gtt} and \eqref{sol_grr}. Let us define the relative error $\nu_{g_{xx}}$ for the metric component $g_{xx}$, describing either $g_{tt}$ or $g_{rr}$, as
\begin{equation}
\nu_{g_{xx}}=\left|\frac{g_{xx}^\text{num}-g_{xx}^\text{fit}}{g_{xx}^\text{num}} \right| \ ,
\end{equation}
where $g_{xx}^\text{num}$ corresponds to the numerical solution obtained via the shooting method and $g_{xx}^\text{fit}$ the analytical fit to the same metric component. In the panels of Fig.\ref{fig:errors}, we provide the plots for the relative errors on both $g_{tt}$ and $g_{rr}$ for all the bosonic star configurations considered. These relative errors are shown to be always smaller than $1\%$, with their average in the range of the radial coordinate $0<x<50$ of the order of $0.01\%$.

\begin{thebibliography}{99}

\bibitem{Crispino:2019yew}
L.~C.~B.~Crispino and D.~Kennefick,
Nature Phys. \textbf{15} (2019) 416.

\bibitem{Luminet:1979nyg}
J.~P.~Luminet,
Astron. Astrophys. \textbf{75} (1979) 228.

\bibitem{Falcke:1999pj}
H.~Falcke, F.~Melia and E.~Agol,
Astrophys. J. Lett. \textbf{528} (2000) L13.


\bibitem{Akiyama:2019cqa}
K.~Akiyama, \textit{et al.} [Event Horizon Telescope],
Astrophys. J. Lett. \textbf{875} (2019) L1.


\bibitem{EHT2022PaperI}
K.~Akiyama, \textit{et al.} [Event Horizon Telescope],  Astrophys. J. Lett. \textbf{930} (2022) L12.

\bibitem{EventHorizonTelescope:2020qrl}
D.~Psaltis \textit{et al.} [Event Horizon Telescope],
Phys. Rev. Lett. \textbf{125} (2020) 141104.

\bibitem{Bozza:2002zj}
V.~Bozza,
Phys. Rev. D \textbf{66} (2002) 103001.

\bibitem{Perlick:2021aok}
V.~Perlick and O.~Y.~Tsupko,
Phys. Rept. \textbf{947} (2022) 1.

\bibitem{Hou:2022gge}
Y.~Hou, P.~Liu, M.~Guo, H.~Yan and B.~Chen,
[arXiv:2203.02755 [gr-qc]].

\bibitem{Cunha:2018acu}
P.~V.~P.~Cunha and C.~A.~R.~Herdeiro,
Gen. Rel. Grav. \textbf{50} (2018) 42.

\bibitem{cunningham}
C. T. Cunningham, J. M. Bardeen,
Astrophysical Journal, \textbf{173} (1972) L137.

\bibitem{Cardoso:2019dte}
V.~Cardoso and R.~Vicente,
Phys. Rev. D \textbf{100} (2019) 084001.

\bibitem{Gralla:2020srx}
S.~E.~Gralla, A.~Lupsasca and D.~P.~Marrone,
Phys. Rev. D \textbf{102} (2020) 124004.

\bibitem{Cardoso:2021sip}
V.~Cardoso, F.~Duque and A.~Foschi,
Phys. Rev. D \textbf{103} (2021) 104044.

\bibitem{Gralla:2019xty}
S.~E.~Gralla, D.~E.~Holz and R.~M.~Wald,
Phys. Rev. D \textbf{100} (2019) 024018.

\bibitem{Vincent:2022fwj}
F.~H.~Vincent, S.~Gralla, A.~Lupsasca and M.~Wielgus,
[arXiv:2206.12066 [astro-ph.HE]].

\bibitem{Lara:2021zth}
G.~Lara, S.~H.~V\"olkel and E.~Barausse,
Phys. Rev. D \textbf{104} (2021) 124041.

\bibitem{Wielgus:2021peu}
M.~Wielgus,
Phys. Rev. D \textbf{104} (2021) 124058.

\bibitem{Vincent:2020dij}
F.~H.~Vincent, \textit{et al.}
Astron. Astrophys. \textbf{646} (2021) A37.

\bibitem{Cardoso:2019rvt}
V.~Cardoso and P.~Pani,
Living Rev. Rel. \textbf{22} (2019) 4.

\bibitem{Barack:2018yly}
L.~Barack, \textit{et al.}
Class. Quant. Grav. \textbf{36} (2019) 143001.

\bibitem{Wielgus:2020uqz}
M.~Wielgus, J.~Horak, F.~Vincent and M.~Abramowicz,
Phys. Rev. D \textbf{102} (2020) 084044.

\bibitem{Tsukamoto:2021caq}
N.~Tsukamoto,
Phys. Rev. D \textbf{104} (2021) 064022.

\bibitem{Olmo:2021piq}
G.~J.~Olmo, D.~Rubiera-Garcia and D.~S.~C.~G\'omez,
Phys. Lett. B \textbf{829} (2022) 137045.

\bibitem{Guerrero:2022qkh}
M.~Guerrero, G.~J.~Olmo, D.~Rubiera-Garcia and D.~G\'omez S\'aez-Chill\'on,
Phys. Rev. D \textbf{105} (2022) 084057.

\bibitem{Tsukamoto:2022vkt}
N.~Tsukamoto,
Phys. Rev. D \textbf{105} (2022) 084036.

\bibitem{Cunha:2017qtt}
P.~V.~P.~Cunha, E.~Berti and C.~A.~R.~Herdeiro,
Phys. Rev. Lett. \textbf{119} (2017) 251102.

\bibitem{Vincent:2015xta}
F.~H.~Vincent, Z.~Meliani, P.~Grandclement, E.~Gourgoulhon and O.~Straub,
Class. Quant. Grav. \textbf{33} (2016) 105015.

\bibitem{Cunha:2017wao}
P.~V.~P.~Cunha, J.~A.~Font, C.~Herdeiro, E.~Radu, N.~Sanchis-Gual and M.~Zilh\~ao,
Phys. Rev. D \textbf{96} (2017) 104040.

\bibitem{Herdeiro:2017fhv}
C.~A.~R.~Herdeiro, A.~M.~Pombo and E.~Radu,
Phys. Lett. B \textbf{773} (2017) 654.

\bibitem{Olivares:2018abq}
H.~Olivares, {\it et al.}
Mon. Not. Roy. Astron. Soc. \textbf{497} (2020) 521.

\bibitem{Guth:2014hsa}
A.~H.~Guth, M.~P.~Hertzberg and C.~Prescod-Weinstein,
Phys. Rev. D \textbf{92} (2015) 103513.

\bibitem{Macedo:2013jja}
C.~F.~B.~Macedo, P.~Pani, V.~Cardoso and L.~C.~B.~Crispino,
Phys. Rev. D \textbf{88} (2013) 064046.

\bibitem{Macedo:2013qea}
C.~F.~B.~Macedo, P.~Pani, V.~Cardoso and L.~C.~B.~Crispino,
Astrophys. J. \textbf{774} (2013) 48.

\bibitem{Brito:2015yfh}
R.~Brito, V.~Cardoso, C.~F.~B.~Macedo, H.~Okawa and C.~Palenzuela,
Phys. Rev. D \textbf{93} (2016) 044045.

\bibitem{Herdeiro:2019mbz}
C.~Herdeiro, I.~Perapechka, E.~Radu and Y.~Shnir,
Phys. Lett. B \textbf{797} (2019) 134845.

\bibitem{Annulli:2020lyc}
L.~Annulli, V.~Cardoso and R.~Vicente,
Phys. Rev. D \textbf{102} (2020) 063022.

\bibitem{Liebling:2012fv}
S.~L.~Liebling and C.~Palenzuela,
Living Rev. Rel. \textbf{15} (2012) 6.

\bibitem{Schunck:2003kk}
F.~E.~Schunck and E.~W.~Mielke,
Class. Quant. Grav. \textbf{20} (2003) R301.

\bibitem{Brito:2015pxa}
R.~Brito, V.~Cardoso, C.~A.~R.~Herdeiro and E.~Radu,
Phys. Lett. B \textbf{752} (2016) 291.

\bibitem{Minamitsuji:2018kof}
M.~Minamitsuji,
Phys. Rev. D \textbf{97} (2018) 104023.

\bibitem{Cardoso:2021ehg}
V.~Cardoso, C.~F.~B.~Macedo, K.~i.~Maeda and H.~Okawa,
Class. Quant. Grav. \textbf{39} (2022) 034001.

\bibitem{Herdeiro:2021lwl}
C.~A.~R.~Herdeiro, A.~M.~Pombo, E.~Radu, P.~Cunha, V.P. and N.~Sanchis-Gual,
JCAP \textbf{04} (2021) 051.

\bibitem{Rosa:2022toh}
J.~L.~Rosa, P.~Garcia, F.~H.~Vincent and V.~Cardoso,
Phys. Rev. D \textbf{106} (2022) 044031.


\bibitem{Visinelli:2021uve}
L.~Visinelli,
Int. J. Mod. Phys. D \textbf{30} (2021) 2130006.

\bibitem{Gleiser:1988rq}
M.~Gleiser,
Phys. Rev. D \textbf{38} (1988), 2376
[erratum: Phys. Rev. D \textbf{39} (1989) no.4, 1257]

\bibitem{Cardoso:2014sna}
V.~Cardoso, L.~C.~B.~Crispino, C.~F.~B.~Macedo, H.~Okawa and P.~Pani,
Phys. Rev. D \textbf{90} (2014) 044069.


\bibitem{Cunha:2015yba}
P.~V.~P.~Cunha, C.~A.~R.~Herdeiro, E.~Radu and H.~F.~Runarsson,
Phys. Rev. Lett. \textbf{115} (2015) 211102.


\bibitem{Gold:2020iql}
R.~Gold,  \textit{et al.}
Astrophys. J. \textbf{897} (2020) 148.

\bibitem{Cunha:2019hzj}
P.~V.~P.~Cunha, N.~A.~Eir\'o, C.~A.~R.~Herdeiro and J.~P.~S.~Lemos,
JCAP \textbf{03} (2020) 035


\bibitem{Peng:2020wun}
J.~Peng, M.~Guo and X.~H.~Feng,
Chin. Phys. C \textbf{45} (2021) 085103.

\bibitem{Wang:2020emr}
X.~Wang, P.~C.~Li, C.~Y.~Zhang and M.~Guo,
Phys. Lett. B \textbf{811} (2020) 135930.

\bibitem{Zeng:2020vsj}
X.~X.~Zeng and H.~Q.~Zhang,
Eur. Phys. J. C \textbf{80} (2020) 1058.

\bibitem{Li:2021riw}
G.~P.~Li and K.~J.~He,
JCAP \textbf{06} (2021) 037.

\bibitem{Chael:2021rjo}
A.~Chael, M.~D.~Johnson and A.~Lupsasca,
Astrophys. J. \textbf{918} (2021) 6.

\bibitem{Cunha:2022gde}
P.~V.~P.~Cunha, C.~Herdeiro, E.~Radu and N.~Sanchis-Gual,
[arXiv:2207.13713 [gr-qc]].



\bibitem{Johnson:2019ljv}
M.~D.~Johnson, \textit{et al.}
Sci. Adv. \textbf{6} (2020) eaaz1310.


\end{thebibliography}
\end{document}